\documentclass[aps,prb,nobalancelastpage,superscriptaddress,twocolumn]{revtex4-2}
\usepackage[utf8]{inputenc}
\usepackage[american,british]{babel}
\usepackage[T1]{fontenc}
\usepackage[pdftex]{graphicx}  
\usepackage{xcolor}
\usepackage{dcolumn}
\usepackage{physics}
\usepackage{tikz}
\usepackage{bm}
\usepackage{amsmath,amsthm,amssymb}
\usepackage{color}
\usepackage{verbatim}
\usepackage[normalem]{ulem}
\usepackage{dsfont}
\usepackage{hyperref}

\usepackage{mathtools}
\hypersetup{
 colorlinks=true,
 linkcolor=blue,
 anchorcolor = blue,
 citecolor = blue,
 filecolor = blue,
 urlcolor = blue
}

\usetikzlibrary{matrix}
\usetikzlibrary{decorations.markings}
\usetikzlibrary{patterns}
\usetikzlibrary{arrows.meta}
\usepackage{graphicx}
\usepackage{subcaption}
\usepackage{caption}
\def \be {\begin{equation}} 
\def \ee {\end{equation}}

\date{}

\usepackage{CircuitTikz}
\usepackage{bbm}
\newcommand{\1}{\mathbbm{1}}

\begin{document}

\title{Dynamical correlation functions of extensive charges after global quantum quenches}
\date{\today}

\author{Riccardo Travaglino} 
\affiliation{SISSA and INFN Sezione di Trieste, via Bonomea 265, 34136 Trieste, Italy}
\author{Katja Klobas} 
\affiliation{School of Physics and Astronomy, University of Birmingham, Edgbaston, Birmingham, B15 2TT, UK}
\author{Bruno Bertini} 
\affiliation{School of Physics and Astronomy, University of Birmingham, Edgbaston, Birmingham, B15 2TT, UK}
\author{Pasquale Calabrese} 
\affiliation{SISSA and INFN Sezione di Trieste, via Bonomea 265, 34136 Trieste, Italy}

\begin{abstract}
We investigate the $n$-time cumulant generating function (or $n$-Full Counting Statistics, $n$-FCS) of extensive $U(1)$ charges following a quantum quench. Exploiting space-time duality we characterise this function when evolving from initial states that are symmetric under the action of the charge. In particular, we show that if the correlations in time are sufficiently weak, e.g.\ the transport is ballistic, the $n$-FCS factorises into a sum of single-time FCS arranged in a time-shell structure. A direct implication of this structure is a drastic simplification of dynamical correlation functions: in the presence of time ordering they only depend on the smallest time entering the correlator. We support these findings with several analytical and numerical tests performed in free and interacting models.
\end{abstract}

\maketitle 

\section{Introduction }

Characterising the properties of out-of-equilibrium many-body quantum systems is one of the central challenges in modern theoretical physics. In the absence of a comprehensive framework comparable to that available for equilibrium settings, progress typically relies on the study of simplified protocols that both capture key aspects of nonequilibrium behaviour and remain analytically tractable.
Arguably the most significant of such protocols is the quantum quench, in which the system is initialised in some low entangled  many-body state $\ket{\Psi_0}$ (usually the ground state of some pre-quench Hamiltonian $H_0$) and let to evolve via a many-body Hamiltonian $H$. 
The study of such protocols has led to the unveiling of fundamental features of quantum evolution, including the linear growth of entanglement~\cite{quench1,quench2,fagotti2008evolution,Calabrese_2012,Calabrese_2012_quantum,alba1,alba2,nahum2017,zhou_nahum_2020,linear_experiment}, local relaxation to thermal or more exotic ensembles~\cite{Deutsch1991,srednicki1,Rigol:2007juv,Caux_2016, Vidmar_2016,ilievski,DAlessio:2015qtq,Gogolin_2016,abanin2019many}, light cone spreading of correlations \cite{Lieb1972,Lauchli_2008,Kliesch2014}, prethermalisation \cite{Fagotti_2014,bertini_2015_prethermalization,Bertini:2015xxd,alba_2017_prethermalization} and generalised hydrodynamics \cite{NESS,NESS2,GHD1,GHD2,ghd_lectures,Alba_2021} to name a few. It is now well established that integrable systems, which harbour ballistic conserved modes~\cite{caux2013,essler2016quench,alba1,alba2,Bertini_2018,Entanglement_GHD_interacting,PIROLI2017362}, exhibit structural differences compared to chaotic models \cite{Deutsch1991,srednicki1,rigol2007,Rigol:2007juv,DAlessio:2015qtq}, with the main distinctions lying in the statistical properties of the spectrum \cite{berrytabor,Bohigas1984spectrum,BORGONOVI20161,he2025statisticalsignaturesintegrablenonintegrable}. 
It can therefore seem puzzling that such distinct classes of models share fundamental features of quench dynamics, most notably the linear growth of entanglement, even though this behaviour arises from entirely different mechanisms in the two settings: ballistic quasiparticle spreading in integrable systems \cite{quench2,alba1}, and the minimal membrane picture in chaotic ones \cite{nahum2017,zhou_nahum_2020}. This highlights that certain aspects of quantum evolution are more universal than the integrable/non-integrable distinction might suggest.

Although originally formulated in Hamiltonian systems, quench protocols have found rich applications in the study of quantum circuits \cite{vonKeyserlingk2018,dualunitary_original,Bertini_2020_OE,piroli2021dualunitary,piroli2021,Potter2022,fisher2023}, where time evolution is implemented through local unitaries, and which can sometimes be interpreted as the time discretisation of integrable or chaotic lattice models. 
In this context, space–time duality~\cite{bertini2026exactly} has emerged as a unifying framework that brings together results from both integrable and chaotic systems under a common perspective. 
The core of this technique relies on the idea to map the complicated non-equilibrium evolution of the system to the stationary properties of a dual system defined in the crossed channel, i.e., in which the roles of space and time are exchanged. This approach has enabled to access several many-body quantities of interest~\cite{bertini2019_minimalmodel,ippoliti2022,PRX,bertini_negativity,bertini2023nonequilibrium,bertini_asymmetric_2024}, which have recently become experimentally accessible~\cite{Greiner2015, Greiner2016, linke2018, greiner2019, elben2020mixed, wei2022quantum, elben2023randomized, rosenberg2024dynamics, joshi2025measuring, yang2026probing}. 
In particular, the two possible viewpoints for linear growth of entanglement are unified by space-time duality, which explains linear growth precisely by the stationarity of the state in the crossed channel. 

While most attention has traditionally focused on the local thermalisation of states, and thus on the analysis of few-site observables, in recent years the focus has been gradually shifted towards studying the statistical properties of extended operators within the system~\cite{klich2009quantum,eisler2013full,eisler2013universality,lovas2017full,najafi2017full,collura2017full,bastianello2018from,calabrese2020full,doyon2023ballistic,oshima2023charge, tartaglia2022real,parez2021quasiparticle,parez2021exact,bertini2023nonequilibrium,bertini_asymmetric_2024,horvath2026full,horvath2026hydrodynamic,travaglino2026space}.
In particular, if the system has a macroscopic conserved charge $Q=\sum_i \hat q_i$, 
 the study of its restriction $Q_A$ to a subsystem $A$  is expected to retrieve some universal features of the quench dynamics, in particular regarding transport within the system. 
In this work, we build on this insight by considering the structure of multi-time dynamical correlation functions of such extended charges in models amenable to the space–time duality approach. Specifically, we show that in the presence of ballistic transport, connected time-ordered correlation functions for quenches from symmetric states in the limit of large subsystem size $\ell_A$ satisfy
\begin{equation} \label{eq:triviality}
  \expval{\mathcal{T}Q_A(t_n) \cdots Q_A(t_2) Q_A(t_1)}_c 
  = \expval*{Q_A^n(\min_it_i)}_c, 
\end{equation}
implying that only the smallest time appearing in the correlator is relevant for the correlation functions. 

The structure of the paper is as follows. In Section~\ref{sec:protocol} we introduce the protocol of interest. In Section~\ref{sec:free} we study the problem in free theories, where we can give a simple quasiparticle explanation to Eq.~\eqref{eq:triviality}. In Section~\ref{sec:SD} we show that this structure can be understood more broadly through space–time duality arguments, which in turn allows us to extend the free-fermionic result to interacting models. In Section~\ref{sec:rule54} we exactly recover Eq.~\eqref{eq:triviality} in the interacting integrable cellular automaton Rule 54, while in Sec.~\ref{sec:TBA} we discuss the emergence of Eq.~\eqref{eq:triviality} in general TBA integrable models. Section~\ref{sec:concl} contains our conclusions, while some of the more technical details of our calculations are relegated to the appendices.

\section{The protocol}
\label{sec:protocol}
We consider lattice Hamiltonians with a global $U(1)$ conserved charge
\begin{equation} 
\label{eq:charge}
   Q = \sum_{j=1}^{L} \hat q_{j}, 
\end{equation}
satisfying $[Q,H]=0$. The system is initialised in a state that is not an eigenstate of the Hamiltonian but is an eigenstate of the charge, ensuring that the subsequent evolution takes place within a fixed-charge sector. We are interested in the multi-time fluctuations of $Q_A$, that is the charge $Q$ restricted to a subsystem $A$. These are characterised by the multi-time full counting statistics
\begin{equation} \label{eq:nfcs_def}
   \mathcal{F}^{(n)}_A(\{t_i,\beta_i\})   
   = \log \Tr\left[\prod_{i=1}^n \mathrm{e}^{i\beta_i Q_A(t_i)} \rho(0)\right]\!,
\end{equation}
where the product should be understood to be implicitly time-ordered from left to right with decreasing time.
Such a generating function gives the connected time-ordered correlation functions by differentiating in the $\beta_i$, 
\begin{equation} \label{correlator}
  \!\!\!\!\!\!\!\expval{Q_A(t_n) \cdots Q_A(t_1)}_c \!\!=\! 
  i^{-n}\partial_{\beta_n} \!\!\!\cdots \partial_{\beta_1}    
  \mathcal{F}^{(n)}_A(\{t_i,\beta_i\})\big |_{\beta_j=0}\!\!\!
\end{equation}
which captures the correlations between charge measurements at different times. As the charge is globally conserved, the non-trivial evolution of $Q_A$ is only due to the charge density flowing through the boundaries of the system, and therefore characterising the correlations gives us insight into transport properties.

Note that the protocol is straightforwardly extended to more general unitary time-evolution that is not necessarily generated by a local time-independent Hamiltonian, such as that given by a quantum circuit where time-evolution is given in discrete steps, $\ket{\psi_{t}}=U_t \ket{\psi_{t-1}}$, with $U_t$ being the time-evolution operator. In this setting a charge $Q$ is conserved if it commutes with the time-evolution operator, i.e.\ $[Q,U_t]=0$ for every $t$. Commonly one considers time-evolution that is periodic in time, in which case we can relax the condition and require that $Q$ is preserved after a finite number of time-steps.

\section{Insight from free theories} 
\label{sec:free}

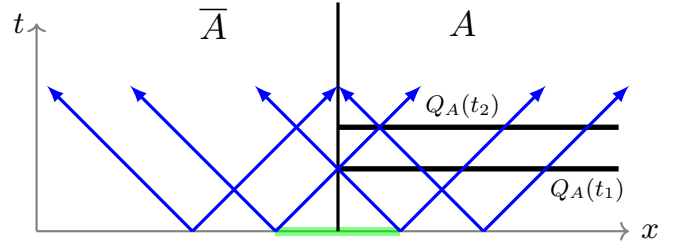
\begin{figure}[t!]
  \centering
  \begin{tikzpicture}[scale=0.55]
      \draw[black,line width=2pt] (7,1.5)--(13.75,1.5) node at (13,1) {$Q_A(t_1)$} ;
      \draw[black,line width=2pt] (7,2.5)--(13.75,2.5) node at (10,3) {$Q_A(t_2)$} ;
      \draw[black,thick,->, gray] (-.25,0) -- (14,0) node[right,scale=1.3, black]{$x$};
      \draw[black,thick,->, gray] (-.25,0) -- (-.25,5) node [left, scale=1.3, black]{$t$};
      \node [scale=1.5] at (4,5){$\overline{A}$};
      \node [scale=1.5] at (10,5){$A$};
        \fill [color=green, opacity=0.5] (5.5,-0.1) -- (8.5,-0.1) -- (8.5,0.1) -- (5.5,0.1) -- cycle;
       \draw[black,very thick, black] (7,0) -- (7,5.5);
       \draw[black,very thick,-latex, blue] (-2+.5+5,0) -- (-2+.5+8.5,3.5);
       \draw[black,very thick,-latex, blue] (-2+.5+5,0) -- (-2+2,3.5);
       \draw[black,very thick,-latex, blue] (.5+5,0) -- (.5+8.5,3.5);
       \draw[black,very thick,-latex, blue] (.5+5,0) -- (2,3.5);
        \draw[black,very thick,-latex, blue] (3+.5+5,0) -- (3+.5+8.5,3.5);
       \draw[black,very thick,-latex, blue] (3+.5+5,0) -- (3+2,3.5);
        \draw[black,very thick,-latex, blue] (5+.5+5,0) -- (5+.5+8.5,3.5);
       \draw[black,very thick,-latex, blue] (5+.5+5,0) -- (5+2,3.5);
  \end{tikzpicture}
  \caption{Quasiparticle-picture explanation of Eq.~\eqref{eq:triviality}. The only quasiparticle pairs contributing to the connected correlator are those emanated from the green region. These correlated pairs will cross all the charge lines once independent of their vertical position.
  \label{fig:qp}}
\end{figure}

It is convenient to begin our discussion thinking about free fermionic systems evolving from initial states that produce correlated fermionic pairs. In this case our main result in Eq.~\eqref{eq:triviality} follows directly from the quasiparticle picture. For simplicity, let us consider
\be
\label{eq:twopoint}
\expval{Q_A(t_2) Q_A(t_1)}_c, \qquad t_2\geq t_1,
\ee
whose quasiparticle structure is depicted in Fig.~\ref{fig:qp}. Because we are interested in a connected correlator, the only pairs contributing to Eq.~\eqref{eq:twopoint} are those crossing all the charge lines. Moreover, since the initial state is an eigenstate of the charge, whenever both correlated quasiparticles from the same pair cross the same charge line the contribution of the charge disappears. The latter fact follows from writing the initial state and the charge in terms of the modes
\be
\ket{\psi_0} = \bigotimes_{k>0} \ket{\psi_{k,-k}}, \qquad  Q = \sum_k \hat q_k ,
\ee
and noting that the eigenstate condition implies 
\be
\!\!(\hat q_k+ \hat q_{-k})\ket{\psi_{k,-k}} = q_{k,-k} \ket{\psi_{k,-k}}, \,\, q_{k,-k}\in\{0,1,2\}\,. 
\ee
Therefore, when the charge operators of both correlated modes act on the initial state the charge becomes a $c$ number and drops out of the correlation. Putting all together, we have that the only quasiparticles contributing to the l.h.s.\ of Eq.~\eqref{eq:twopoint} are those originating in the interval $x\in[-t_1,t_1]$ (green interval in the figure). All these pairs cross the second charge line independently of $t_2\geq t_1$, i.e., the result is $t_2$-independent and we can set $t_2=t_1$. The same reasoning is immediately generalised to find Eq.~\eqref{eq:triviality}.

This heuristic argument can be formalised using the operatorial quasiparticle picture developed in Ref.~\cite{Bertini_2018} (see also Refs.~\cite{rottoli2024entanglementhamiltoniansquasiparticlepicture,travaglino2024, molly}), which allows for a precise characterisation of the entanglement Hamiltonian and other aspects of non-equilibrium dynamics~\cite{travaglino2025,Travaglino2025measurements,travaglino2026dissipative}. In this approach, the quench dynamics at the ballistic scale is fully determined by the group velocity $v_k = \partial_k\varepsilon_k$ and the occupation functions $\theta_k$. As shown in App.~\ref{appA}, this framework allows us to obtain the $n$-time FCS by considering the evolution of coarse-grained degrees of freedom on which $Q_A$ acts simply. In the regime $t_n\ll \ell$, for symmetric initial states, we arrive at
\begin{equation} \label{eq:Fdimerfinal}
  \begin{aligned}
    \mathcal{F}^{(n)}_A(\{t_i,\beta_i\})&=i q_0\ell_A \bar{\beta}(0)\\
    &+\sum_{\sigma=\pm 1} \int\frac{{\rm d}k}{2\pi} 
    v_k \int_0^t dt'
    \mathcal{L}(k,\sigma\bar{\beta}(t^{\prime})), 
  \end{aligned}
\end{equation}
where $q_0$ is the expectation value of the density of the charge in the initial state, $\sigma\in\{+1,-1\}$ denotes the contribution from either left or right edge of the subsystem, 
$\bar{\beta}(t)$ is the sum of $\beta_j$ with $t_j>t$
\begin{equation}\label{eq:defBarBeta}
  \bar{\beta}(t)=\sum_{j=1}^n \beta_j \,\Theta(t_j-t),\qquad 
  \Theta(x)=\begin{cases}1, \quad& x>0,\\
    0,& x\le 0,
  \end{cases}
\end{equation}
and we have introduced the function
\begin{equation}\label{Ldefinition}
    \mathcal{L}(k,{\beta})={\rm sgn}(v_k)
    \log\left[\frac{\theta_k}{\mathrm{e}^{-i\ {\rm sgn}(v_k){\beta}}}+(1-\theta_k)\right].
\end{equation}
By generalising the result of FCS to $n$-FCS we have thus obtained a \emph{time-shell} structure given by $\bar{\beta}(t)$ in Eq.~\eqref{eq:defBarBeta}, and so we can integrate separately over every interval $[t_{j},t_{j+1}]$ to rewrite the $n$-FCS as a sum of contributions of each time-shell
\begin{equation}\label{eq:additivity}
  \mathcal{F}^{(n)}_A(\{t_i,\beta_i\})=i q_0\ell_A \bar{\beta}_0
  +\sum_{j=0}^{n-1} \Delta t_{j} \Gamma(\bar{\beta}_j),
\end{equation}
where we have isolated the contribution
\begin{equation}
  \Gamma({\beta})= \int\frac{{\rm d}k}{2\pi} v_k
  \left[\mathcal{L}(k,{\beta})
  +\mathcal{L}(k,-{\beta})\right],
\end{equation}
and introduced the shorthand notation $\Delta t_j=t_{j+1}-t_j$, with $t_0\equiv 0$, and $\bar{\beta}_j\equiv\bar{\beta}(t_j)=\sum_{k>j} \beta_k$. Since we have
\begin{equation}
  \left.\frac{\partial \bar{\beta}_j}{\partial \beta_i}\right|_{i\le j} = 0,
\end{equation}
the only non-trivial contribution in the evaluation of the correlator in Eq.~\eqref{correlator} comes from the first time-shell. This yields 
\begin{align}
 \expval{Q_A(t_n) \cdots Q_A(t_1)}_c &= 
  i^{-n}\partial_{\beta_n} \cdots \partial_{\beta_1}    
  \mathcal{F}^{(n)}_A(\{t_i,\beta_i\})\big  |_{\beta_j=0} \notag\\
  &= \Delta t_{0}   i^{-n}\partial_{\beta_n} \cdots \partial_{\beta_1} \Gamma(\bar{\beta}_0)\big  |_{\beta_j=0}\notag\\
  &= i^{-n}\partial_{\beta_n} \cdots \partial_{\beta_1} \mathcal{F}^{(1)}_A(t_1,\bar{\beta}_0)\big  |_{\beta_j=0}\notag \\
  &=\expval{Q^n_A(t_1)}_c.
\end{align}
Namely, Eq.~\eqref{eq:triviality}.

To confirm this result numerically we consider the XX chain
\begin{equation}
    H_{XX} = \sum_i \sigma^x_i \sigma^x_{i+1} + \sigma^y_i \sigma^y_{i+1},
\end{equation}
which can be mapped into a free fermionic system (the tight binding model) through a Jordan-Wigner transformation. 

\begin{figure}
    \centering
    \includegraphics[width=\linewidth]{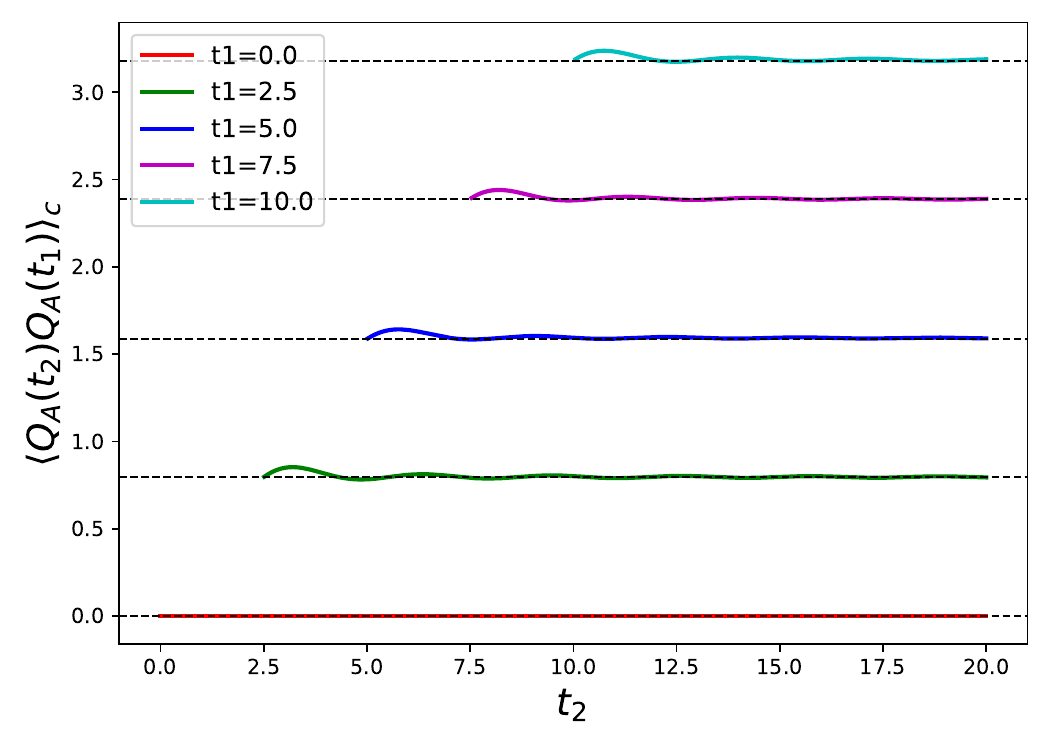}
    \caption{Connected two point function vs $t_2$ after a quench from the fermionic N\'eel state in the tight binding model for a subsystem of $\ell_A =40$ sites. Dashed lines represent $\expval{Q_A^2(t_1)}_c$. The plot shows that, to a very good approximation, the correlator only depends on the earliest time $t_1$.
    \label{fig:2ptfunction}}
\end{figure}

\begin{figure}
  \centering
  \includegraphics[width=\linewidth]{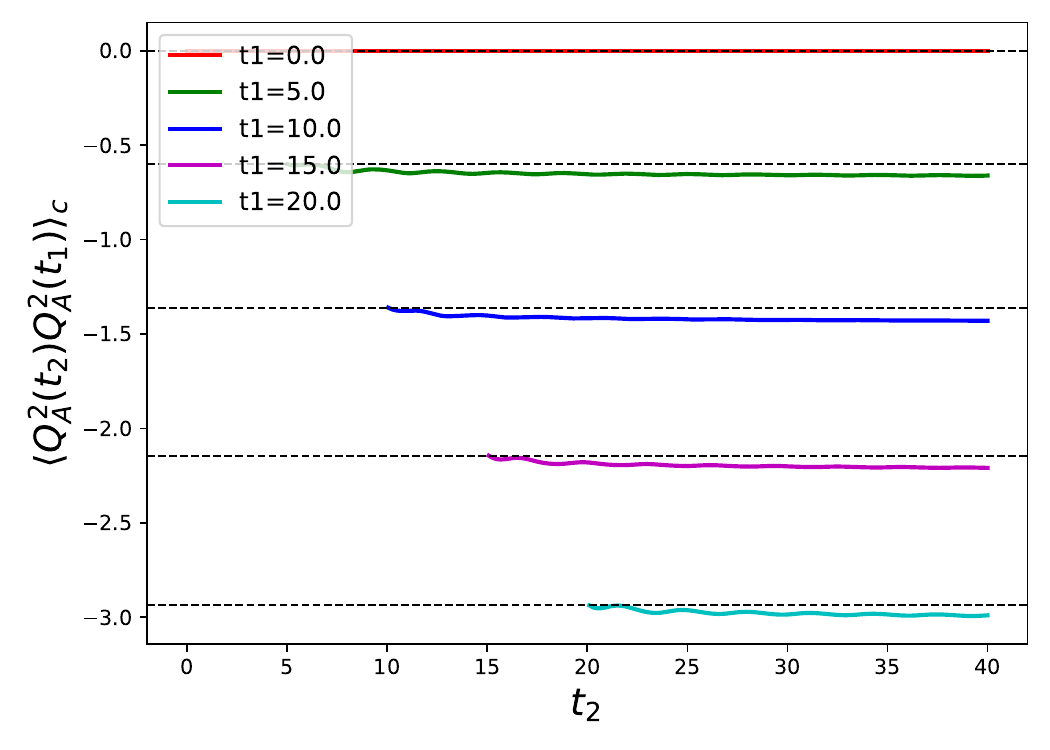}
  \caption{Connected 4-point function  $\expval{Q^2_A(t_2) Q^2_A(t_1)}_c$ vs $t_2$ after a quench from the fermionic N\'eel state in the tight binding model for subsystem of $\ell_A =80$ sites. Compared to the 2-point function we observe a small offset. This reflects subleading contributions that are present also in the single-time 4-point function $\expval{Q^4_A(t_1)}_c$. As shown in Appendix~\ref{appC}, these corrections are reduced when increasing the system size.
  \label{fig:3pointfunction}}
\end{figure}

In fermionic language, numerical evaluations can be simply implemented by correlation matrix techniques, which allow to easily reach system sizes of several hundreds of sites. In a nutshell, these techniques rely on the fact that the free evolution preserves Gaussianity, and therefore if the initial state is Gaussian the correlation functions at all times are fully determined by the two-time correlation matrices $C_{xy}(t_2,t_2) = \expval{c_x^\dag(t_2) c_y(t_1)}$ and $\tilde C_{xy}(t_2,t_1) = \expval{c_x(t_2) c^\dag_y(t_1)}$ by Wick theorem. In many quenches of interest, the elements of these matrices can be evaluated exactly in terms of Bessel functions, allowing for a complete analytical characterisation \cite{Eisler_2007,Peschel_2009}. This insight allows to obtain the results shown in Figs.~\ref{fig:2ptfunction} and \ref{fig:3pointfunction}, which contain the 2 and 4-point functions respectively for quenches from the fermionic N\'eel state  $\ket{1010\dots}$. The results do indeed confirm that the full time dependence at times smaller than the size of $A$ is asymptotically fixed by the first time $t_1$, through the relations
\begin{equation} \label{eq:correlatorsnumerics}
  \begin{aligned}
    \expval{Q_A(t_2)Q_A(t_1)}_c &=   \expval{Q^2_A(t_1)}_c, \\
    \expval{Q^2_A(t_2) Q^2_A(t_1)}_c &=   \expval{Q^4_A(t_1)}_c,
  \end{aligned}
\end{equation}
in agreement with Eq.~\eqref{eq:triviality}. We have also checked that the three point function $\expval{Q_A(t_3) Q_A(t_2) Q_A(t_1)}_c$ vanishes for all choices of $t_3>t_2>t_1$. This is in line with our expectations since $\expval*{Q^3_A(t_1)}_c=0$ in quenches from the Néel state. In the following section, we show that Eq.~\eqref{eq:triviality} admits a direct interpretation in terms of space-time duality picture, which allows us to extend it to interacting examples. 

\section{Space-time duality} 
\label{sec:SD}
\begin{figure*}[t!]
  \centering
  \begin{tikzpicture}[scale=0.65]
    \begin{scope}
      \fill [cyan!12] (1,0) rectangle (4,5);
      \fill [cyan!12] (8,0) rectangle (11,5);
      \fill [orange!20] (4,0) rectangle (8,5);

      \draw[black,thick,->] (1,0) -- (11,0) node[right,scale=1.3]{$x$};
      \draw[black,thick,->] (1,0) -- (1,5) node [left, scale=1.3]{$t$};

      \draw[black,line width=1.5pt] (4,1.5) -- (8,1.5) node[right]{$\mathrm{e}^{i\beta_1 Q_A}$};
      \draw[black,line width=1.5pt] (4,3.0) -- (8,3.0) node[right]{$\mathrm{e}^{i\beta_2 Q_A}$};
      \draw[black,line width=1.5pt] (4,4.5) -- (8,4.5) node[right]{$\mathrm{e}^{i\beta_3 Q_A}$};

      \node [scale=1.5] at (2.5,5.5){$\overline{A}$};
      \node [scale=1.5] at (9.5,5.5){$\overline{A}$};
      \node [scale=1.5] at (6,5.5){$A$};

      \draw (3.8,0.05) -- (3.8,1.45) node[midway,xshift=-1em]{$\tau$};
      \draw (3.8,1.55) -- (3.8,2.95) node[midway,xshift=-1em]{$\tau$};
      \draw (3.8,3.05) -- (3.8,4.45) node[midway,xshift=-1em]{$\tau$};
    \end{scope}

    \draw[line width=2.5pt,->,>=stealth,black!70] (12,2.5) -- (14.5,2.5);

    \begin{scope}[xshift=15cm]
      \fill [cyan!12] (1,0) rectangle (4,5);
      \fill [cyan!12] (8,0) rectangle (11,5);
      \fill [orange!20] (4,0) rectangle (8,5);

      \draw[black,thick,->] (1,0) -- (11,0) node[right,scale=1.3]{$x$};
      \draw[black,thick,->] (1,0) -- (1,5) node [left, scale=1.3]{$t$};

      \draw[black,line width=1.5pt] (8,1.5) -- (8,0)   node[midway,right]{$\mathrm{e}^{i\sum_i\beta_i \tilde Q_A}$};
      \draw[black,line width=1.5pt] (8,3.0) -- (8,1.5) node[midway,right]{$\mathrm{e}^{i\sum_{i>1}\beta_i\tilde Q_A}$};
      \draw[black,line width=1.5pt] (8,4.5) -- (8,3.0) node[midway,right]{$\mathrm{e}^{i\sum_{i>2}\beta_i \tilde Q_A}$};
      \draw[black,line width=1.5pt] (4,4.5) -- (4,0);

      \draw[black,dashed,line width=1pt] (4,3.0) -- (8,3.0);
      \draw[black,dashed,line width=1pt] (4,1.5) -- (8,1.5);

      \node [scale=1.5] at (2.5,5.5){$\overline{A}$};
      \node [scale=1.5] at (9.5,5.5){$\overline{A}$};
      \node [scale=1.5] at (6,5.5){$A$};

      \draw (3.8,0.05) -- (3.8,1.45) node[midway,xshift=-1em]{$\tau$};
      \draw (3.8,1.55) -- (3.8,2.95) node[midway,xshift=-1em]{$\tau$};
      \draw (3.8,3.05) -- (3.8,4.45) node[midway,xshift=-1em]{$\tau$};

      \draw [black,line width=1.5pt] (4,0) -- (8,0) node[midway,yshift=-1em]{$\mathrm{e}^{i\sum_i \beta_i Q_0}$};
    \end{scope}
  \end{tikzpicture}
    \caption{ Deformation of charge lines at times $n\tau$ to vertical insertions of the dual charge $\tilde Q_A$, which represent insertions of charge lines in the dual theory. The contribution of the initial state is trivial for symmetric initial states, hence $Q_0$ is just a number, and all the relevant physics takes place at the boundaries of $A$. }
  \label{fig:chargedeformation}
\end{figure*}
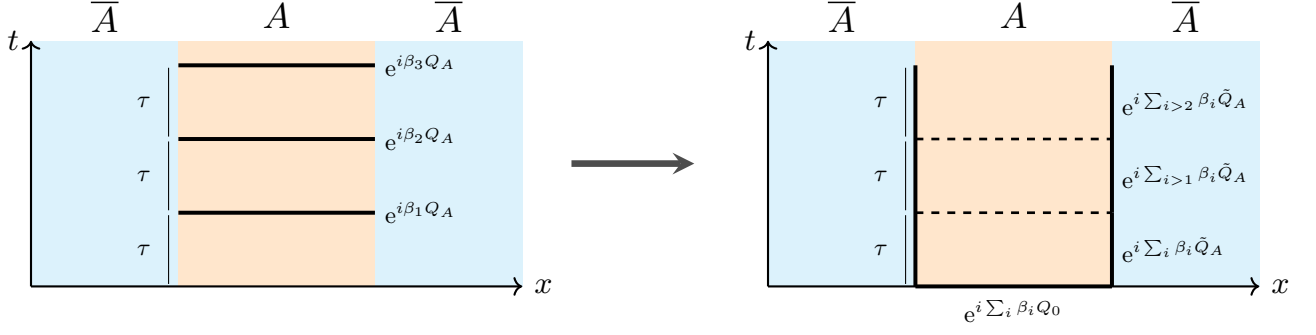

The main idea of space-time duality is to map complicated non-equilibrium properties of a given theory to equilibrium properties of the theory in which the roles of space and time are swapped~\cite{PRX,bertini_negativity,bertini2023nonequilibrium,bertini_asymmetric_2024,bertini2026exactly,travaglino2026space}. 

More concretely, considering the exponential of $\mathcal{F}^{(n)}_A(\{t_i,\beta_i\})$ (cf.~Eq.~\eqref{eq:nfcs_def}) one can deform the charge lines as shown in Fig.~\ref{fig:chargedeformation}, where the vertical lines correspond to time ordered exponentials of the integrated current~\cite{BFT,horvath2024full,DelVecchio2024,horvath2026full,horvath2026hydrodynamic} while the horizontal one is trivial for initial states that are symmetric under the action of the charge. This transformation is a direct consequence of the continuity equation, which makes the line operator $\mathrm{e}^{i\beta_j Q_A}$ topological. 

The right panel of Fig.~\ref{fig:chargedeformation} can now be contracted in space, i.e., from left to right. In this case, for $\ell_A\gg t$, one obtains
\be
\label{eq:circuitFCS}
\begin{aligned}
  \mathrm{e}^{\mathcal{F}^{(n)}_A(\{t_i,\beta_i\})} &= \mathrm{e}^{i Q_0\sum_i \beta_i}  \\
     &\times\prod_{\sigma = \pm 1} \operatorname{Tr}[\tilde{\rho}_{\rm st} \mathrm{e}^{i \sigma \beta_1 \tilde Q_{t_1}} \cdots \mathrm{e}^{i \sigma \beta_n \tilde Q_{t_n}}],
\end{aligned}    
\ee
where $\tilde{\rho}_{\rm st}$ is the product of left and right fixed points of the space-evolution super-operator~\cite{PRX, bertini2023nonequilibrium,bertini_asymmetric_2024, travaglino2026space} and $\tilde Q_t$ is the charge of the system on the time lattice of length $t$. For instance, for a system in discrete time one has~\cite{bertini2023nonequilibrium,bertini_asymmetric_2024}
\begin{equation} 
\label{eq:charge_crossed_channel}
  \tilde Q_t = \sum_{j=1}^{t} \left(\hat q_{2j-1} - \hat{q}_{2j}^T\right), 
\end{equation}
where $\hat q_{j}$ is the charge density in Eq.~\eqref{eq:charge}. In writing Eq.~\eqref{eq:circuitFCS} we assumed that the system has a maximal velocity (e.g.\ the Lieb Robinson bound applies) and the two edges of the interval $A$ become disconnected for sufficiently large $\ell_A$.

Eq.~\eqref{eq:circuitFCS} is an exact rewriting and, as such, is too general to yield any strong conclusions. To make progress we assume that  --- in the swapped channel --- the quantities that are \emph{extensive in time} are additive at leading order. In this case the $n$-FCS further factorises into contributions from different time-shells, if all the time-intervals $\Delta t_i$ are large compared to microscopic space-time scales and we have 
\begin{equation} 
\label{eq:timeshellSD}
  \begin{aligned}
    \mathcal{F}^{(n)}_A(\{t_i,\beta_i\}) &\approx iQ_0 \sum_j\beta_j  \\
    &+\sum_{\sigma=\pm} \sum_j \log\operatorname{Tr}
    \left(\tilde{\rho}_{\rm st}
    \mathrm{e}^{i \sigma \bar \beta_j \tilde{Q}_{[t_j,t_{j+1}]} }\right),
  \end{aligned}
\end{equation}
where we introduced 
\begin{align}
      \tilde Q_{[t_i,t_{i+1}]} &= \sum_{j=t_i}^{t_{i+1}} \left(\hat q_{2j-1} - \hat{q}_{2j}^T\right).
\end{align}
This form has the key feature of Eq.~\eqref{eq:additivity}: $\beta_j$ appears only in the contributions coming from time-shells $[t_i,t_{i+1}]$ with $i<j$. It is therefore clear that $n$-point correlation function will preserve the property in Eq.~\eqref{eq:triviality} whenever the above approximation is justified. Moreover, each term appearing in the sum can be evaluated separately through the techniques of Ref. \cite{bertini2023nonequilibrium}.

The additivity of time-extensive variables seems like a natural assumption, since this holds for expectation values of extensive variables in equilibrium ensembles. We remark, however, that the \emph{dual} stationary state  $\tilde{\rho}_{\rm st}$ is not guaranteed to have these properties, and the linearity of extensive quantities needs to be independently justified. In particular, one needs to require that the correlations in $\tilde{\rho}_{\rm st}$ are short-range (in time). 

To make a more quantitative statement we consider the case of two charge insertions at values $\tau$, $2\tau$ and a single edge. In this case Eq.~\eqref{eq:circuitFCS} can be expressed as 
\be
     \mathcal{F}^{(n)}_A(\{t_i,\beta_i\})=i Q_0\sum_i\beta_i + \sum_{\sigma=\pm} \log\operatorname{Tr}[\tilde{\rho}_{\rm st} \mathrm{e}^{i  \sigma \tilde Q}],  
\ee     
with 
\be     
   \tilde Q = (\beta_1 + \beta_2) \tilde Q_{[0,\tau]} + \beta_2 \tilde Q_{[\tau,2\tau]}.
\ee
Importantly, $\tilde Q_{[0,\tau]}$ and $\tilde Q_{[\tau,2\tau]}$ commute, since they have different supports, implying that the separation condition which can lead to Eq.~\eqref{eq:timeshellSD} originates purely from a cumulant expansion.
Assuming that the correlation functions of the dual charge operators in the different time shells is sufficiently weak compared to the  correlators within each shell, it is simple to see that Eq.~\eqref{eq:circuitFCS} can be approximated as 
\begin{align}
  \mkern-20mu
  \mathcal{F}^{(n)}_A(\{t_i,\beta_i\}) &\approx i Q_0\sum_i\beta_i + \sum_{\sigma=\pm}  
  \log\operatorname{Tr}[\tilde{\rho}_{\rm st} \mathrm{e}^{i \sigma \beta_2 \tilde Q_{[\tau,2\tau]}}]
  \nonumber \\
& + \sum_{\sigma=\pm}  \log\operatorname{Tr}[\tilde{\rho}_{\rm st}
  \mathrm{e}^{i \sigma (\beta_1 + \beta_2) \tilde Q_{[0,\tau]}}] + \Delta_\tau     
 \label{eq:witherror}\mkern-20mu
\end{align}
 where the correcting factor is given by a sum over all cross-cumulants  $\expval*{Q^n_{[0,\tau]} Q^m_{[\tau,2\tau]}}_c^{\tilde\rho_{\rm st}}$. 
Assuming that higher order correlators are subleading with respect to the two-point function this gives
\begin{equation}
  \Delta_\tau \propto  \expval{Q_{[0,\tau]} Q_{[\tau,2\tau]}}_c^{\tilde\rho_{\rm st}} + \rm{h.c.} 
\end{equation}
Since the main terms in Eq.~\eqref{eq:witherror} are linear in $\tau$ in models with ballistic transport, our approximation is asymptotically good as long as the correction $\Delta_\tau$ is sublinear. In a hydrodynamic regime, in which all sums can be approximated by integrals, this can be expanded as
\begin{equation}
    \expval*{\tilde Q_{[0,\tau]} \tilde Q_{[\tau,2\tau]}}_c^{\tilde\rho_{\rm st}} 
    = \int\limits_0^\tau \! dt_1 \!\int\limits_0^\tau \! dt_2 
    \expval{\tilde q(t_1) \tilde q(\tau+t_2)}_c^{\tilde \rho_{\rm st}}
\end{equation}
where $\tilde q(t)$ is the local density. In particular, the integral gives a sublinear term in $\tau$ as long as the local correlator satisfies the clustering property 
\be
\label{eq:sufficientcondition}
 |t-t'| \expval{\tilde q(t) \tilde q(t')}_c^{\tilde \rho_{\rm st}}  \to 0,
 \ee 
 for large time separations. Note that this condition is sufficient but not necessary, as one can see by observing that $\expval{\tilde q(t) \tilde q(t')}_c^{\tilde \rho_{\rm st}} = |t-t'| ^{-1}$ makes the integral linear in $\tau$, but we will consider it as it admits a more direct physical interpretation. At this point, we note that the charge in the crossed channel is dual to the current operator in the regular channel. Therefore, Eq.~\eqref{eq:sufficientcondition} is equivalent to the requirement 
\be
|t-t'| \expval{j(x,t) j(x,t')}_c  \to 0, 
\ee
which was considered in Ref.~\cite{horvath2024full,horvath2026full}. Hence, we conclude that the range of validity of our results is the same as the ones of those references, i.e., they are expected to hold in theories with ballistic transport.  

We stress that strictly speaking the origin of the peculiar result in Eq.~\eqref{eq:triviality} is ballistic transport, not integrability, therefore one can expect it to hold also in chaotic systems with ballistic propagation of charge such as charged dual unitary circuits~\cite{foligno2025nonequilibrium}.

\section{Rule 54}
\label{sec:rule54}

An interacting integrable case where Eq.~\eqref{eq:timeshellSD} can be shown to hold exactly is \emph{Rule 54}, a cellular automaton~\cite{bobenko1993two}, which in recent years has emerged as a ``minimal'' interacting integrable model showing effects of quasi-particle interactions~\cite{friedman2019integrable,gombor2024integrable} but retaining exact solvability~\cite{buca2021rule54}. In particular for us it is important that there exists a quench from an integrable initial state, which one can solve and obtain exact expressions for the asymptotic rates of growth of R\'enyi entanglement entropies, as well as more general charged moments~\cite{klobas2021exact,klobas2021exactrelaxation,klobas2021entanglement,klobas2024nonequilibrium}.

The model is defined on a chain of $L$ qubits with time-evolution given in discrete time-steps,
\begin{equation}
  \ket{\psi_t}=\mathbb{U}_{t}\ket{\psi_{t-1}},\qquad
  \mathbb{U}_t=\smashoperator{\prod_{\substack{1\le x\le L\\ x \equiv t \pmod{2}}}} 
  U_{x-1,x,x+1},
\end{equation}
where $U_{x-1,x,x+1}$ is a \emph{local} time-evolution operator $U$ applied to three neighbouring sites centred around $x$,
\begin{equation}
  U_{x-1,x,x+1}=\1^{\otimes x-2}\otimes U \otimes \1^{\otimes L-x-1}.
\end{equation}
The local operator $U$ updates the middle site according to a deterministic binary rule 
$\chi: \mathbb{Z}_2^{\times 3} \to \mathbb{Z}_2$,
\begin{equation}
  \chi(s_1,s_2,s_3)\equiv s_1+s_2+s_3+s_1 s_3\pmod{2},
\end{equation}
so that the matrix elements of $U$ are 
\begin{equation}
  \mel{s_1 s_2 s_3}{U}{s_1^{\prime} s_2^{\prime} s_3^{\prime}}
  =\delta_{s_1,s_1^{\prime}}
  \delta_{s_3,s_3^{\prime}}
  \delta_{s_2,\chi(s_1^{\prime},s_2^{\prime},s_3^{\prime})}.
\end{equation}
The induced dynamics can be understood in terms of stable quasi-particle excitations that move with fixed velocities $\pm 1$ and undergo pairwise scattering.
The resulting dynamics is integrable, and therefore exhibits a large number of local conserved charges, and we will focus on $Q^{-}$, the net quasi-particle current (i.e.\ the difference between the numbers of left and right movers)
\begin{equation}
  \begin{aligned}
    Q^{-}&=\sum_{j=1}^{L}(-1)^j q^{-}_{j,j+1},\\
    q^{-}_{j,j+1}&=\frac{1}{4}\left(1-\sigma_{j}^z\right)\left(1-\sigma_{j+1}^{z}\right).
  \end{aligned}
\end{equation}
The reason to chose the charge $Q^{-}$ is that the \emph{solvable} initial state $\ket{\Psi_0}$ (that is, the initial state for which the quench can be solved exactly) is its eigenstate
\begin{equation}
  q_{j,j+1}^{-} \ket{\Psi_0}=0,\  \ket{\Psi_0}=
  \left[\sqrt{1-\vartheta}\ket{00}+\sqrt{\vartheta} \ket{01}\right]^{\otimes \frac{L}{2}}
  \mkern-24mu.\ 
\end{equation}
Here $0<\vartheta<1$ is a free parameter that coincides with the values of both the filling functions of the corresponding stationary GGE reached at long times~\cite{klobas2021exactrelaxation}.

As we show in App.~\ref{sec:R54App}, in the regime $|A|>2t$ the $n$-FCS factorises into two contributions coming from the edges of the subsystem
\begin{equation}
  \mathcal{F}_{A}^{(n)}(\{t_i,\beta_i\})=
  \mathcal{F}_{A,+}^{(n)}(\{t_i,\beta_i\})
  +\mathcal{F}_{A,-}^{(n)}(\{t_i,\beta_i\}),
\end{equation}
and these two are given as matrix products
\begin{equation}\label{eq:R54altFpm}
  \mkern-16mu
  \begin{aligned}
    \mathcal{F}_{A,\pm}^{(n)}=
    \log 
    \bra{T}
    &\left(M_{\pm}(\Pi_{n-1})\right)^{\frac{\Delta t_{n-1}}{2}}
    \left(M_{\pm}(\Pi_{n-2})\right)^{\frac{\Delta t_{n-2}}{2}}\\
    &\cdots
    \left(M_{\pm}(\Pi_{0})\right)^{\frac{\Delta t_{0}}{2}}
    \ket{B_{\pm}},
  \end{aligned}
  \mkern-16mu
\end{equation}
where we have introduced $\Pi_j$ as the shorthand for
\begin{equation}
  \Pi_j=\prod_{l=j+1}^{n} \mathrm{e}^{i\beta_l}.
\end{equation}
Here $M_{\pm}$ are $3\times 3$ matrices, and $\bra{T}$, $\ket{B_{\pm}}$ $3$-dimensional vectors, explicitly given by
\begin{equation}
  M_{+}(\Pi)=
  \begin{bmatrix}
    (1-\vartheta)^2 & \Pi & 1-\vartheta \\
    \vartheta(1-\vartheta) & 0 & \vartheta \\
    \Pi \vartheta & 0 & 0
  \end{bmatrix}\mkern-6mu,\ 
  \ket{B_+}=\begin{bmatrix} 1-\vartheta \\ \vartheta \\ 0 \end{bmatrix}\mkern-6mu,
\end{equation}
while the corresponding matrices from the other edge are
\begin{equation}
  M_{-}(\Pi)=
  \begin{bmatrix}
    (1-\vartheta)^2 & 1 & 1-\vartheta \\
    \Pi^{-1}\vartheta(1-\vartheta) & 0 & \Pi^{-1}\vartheta \\
    \Pi^{-1} \vartheta & 0 & 0
  \end{bmatrix}\mkern-6mu,\ 
  \ket{B_{-}}=\begin{bmatrix} 0 \\ 0 \\ 1 \end{bmatrix}\mkern-6mu,
\end{equation}
and the top vector is given by
\begin{equation}
  \bra{T}=\begin{bmatrix} 1 & 1 & 1 \end{bmatrix}.
\end{equation}
This form immediately suggests that in the limit of large $\Delta t_j$ the $n$-FCS reduces to a sum over contributions from different time-shells as in Eq.~\eqref{eq:timeshellSD}. Indeed, for large $\Delta t_j$ Eq.~\eqref{eq:R54altFpm} can be approximated by projecting to (left and right) leading eigenvectors and we have
\begin{equation}\label{eq:finalEqR54}
  \mathcal{F}_{A,\pm}=
  \sum_{j=0}^{n-1}\frac{\Delta t_j}{2} \log \lambda_{\pm}(\bar{\beta}_j)
  + \mathcal{O}\left((n+1)(\Delta t_j)^0\right),
\end{equation}
where 
\begin{equation}
  \bar{\beta}_j=\sum_{l=j+1}^n \beta_j,
\end{equation}
and $\lambda_{\pm}(\beta)$ is the leading eigenvalue of $M_{\pm}(\mathrm{e}^{i \beta})$ given by the largest (in magnitude) solution of the following algebraic equation
\begin{equation}
  \lambda_{\pm}(\beta)=
  \left(1-\vartheta+\frac{\vartheta \mathrm{e}^{\pm i\beta}}{\lambda_{\pm}}\right)^2.
\end{equation}
The leading-order correction in Eq.~\eqref{eq:finalEqR54} comes from overlaps between left and right leading eigenvectors of $M_{\pm}$ evaluated at neighbouring $\Pi_j$ and those with boundary vectors, while the corrections of subleading eigenvalues are exponentially small in $\Delta t_j$.  We note that the expression holds also for $n=1$, in which case we recover the known expression for the FCS from Refs.~\cite{bertini2023nonequilibrium,klobas2024nonequilibrium}.

\section{Bethe-ansatz Integrable models}
\label{sec:TBA}

Let us now consider the case of general Bethe-ansatz integrable models, where, following Refs.~\cite{PRX, bertini2023nonequilibrium,bertini_asymmetric_2024, travaglino2026space} one can conjecture a closed form  expression for the r.h.s.\ of Eq.~\eqref{eq:timeshellSD} and then check its validity against numerical results.  

The starting point is to recall that the thermodynamics of integrable systems is described exactly by the  thermodynamic Bethe ansatz (TBA)~\cite{yangyang,Takahashi_1999, Zamolodchikov1989}. In this framework one characterises collections of similar eigenstates (macrostates) through the total density of available modes $\rho^i_{\rm t}(\mu)$, and the occupation functions $\vartheta^i(\mu)$ of the quasiparticles, where $i$ labels quasiparticle species. These quantities are related through the Bethe-Takahashi equations as follows
\begin{equation}
    \rho_t^i = \frac{p_j'(\mu)}{2\pi} - \sum_k (K_{ik}* \theta^k\rho_t^k)(\mu).
\end{equation}
Here, $K_{ij}(\mu)$ is the scattering kernel of the theory, which encodes the model dependence, and $*$ represents convolution. The rapidity variable $\mu$ is introduced as a parametrisation of the single particle dispersion relation $(p(\mu),\varepsilon(\mu))$. The equations describe excitations propagating at a velocity $v^{dr}(\mu)$, which is renormalised by the effect of the interactions ~\cite{bonnes2014}, through
 \begin{equation}
     \frac{\varepsilon_j'(\mu)}{2\pi} = v^{dr}_j \rho_t^j +  (K_{ik}* \theta^k  v^{dr}_k \rho_t^k)(\mu).
 \end{equation}
In these systems the terms in Eq.~\eqref{eq:timeshellSD} can be directly computed through the space-time swap of the known stationary expressions
\be
\log\operatorname{Tr}
    \left(\rho_{\rm st}
    \mathrm{e}^{i  \beta Q_A} \right) = \ell_A \int \frac{d\mu}{2\pi} p'(\mu) K(\mu,\beta),
\label{eq:equilibriumquantity}
\ee 
where we introduced
\begin{equation}
     \begin{split}
      \label{eq:equilibriumresult}
 &\mathcal{K}(\mu,\beta) =
    \log\left[\frac{\theta(\mu)}{ x(\mu)}+(1-\theta(\mu))\right], \\
   & \log x =  -i\beta q+ \int \frac{d\mu'}{2\pi} K(\mu',\mu) \mathcal{L}(\mu',\bar{\beta}(t)), 
     \end{split}
 \end{equation}
Note that we have assumed the presence of a single particle species, which carries a bare charge $q$; in the case of more general theories it would be simply enough to sum over the different quasiparticles which emerge in Bethe ansatz description.

To find an expression for Eq.~\eqref{eq:timeshellSD} we note that each term appearing in the sum corresponds to Eq.~\eqref{eq:equilibriumquantity} upon an exchange of space and time, with different parameters $\beta_j$ depending on the time shell. Recalling that in TBA integrable models exchanging space and time is equivalent to exchanging energy and momentum of the quasiparticles one can then make the following conjecture
\be
\!\!\!\operatorname{Tr}[\tilde{\rho}_{\rm st} \mathrm{e}^{i \bar \beta_j \tilde{Q}_{[t_j,t_{j+1}]}} ] = (t_{j+1}-t_j) \int \frac{{\rm d}\lambda}{2\pi} \varepsilon'(\mu)\mathcal{L}(\mu,\bar{\beta}_j), 
\label{eq:nonequilibriumquantity}
\ee 
where we introduced
 \begin{equation}
     \begin{split}
     \label{eq:interacting}
 &\mathcal{L}(\mu,\bar{\beta}_j) =s(\mu)
    \log\left[\frac{\theta(\mu)}{ y^{s(\mu)}}+(1-\theta(\mu))\right], \\
   & \log y =  -i q\bar{\beta}_j+ \int \frac{d\mu'}{2\pi} K(\mu',\mu) \mathcal{L}(\mu',\bar{\beta}_j), 
     \end{split}
 \end{equation}
 and factors $s(\mu)= {\rm sgn}(\varepsilon'(\mu))$ are added following the discussion of \cite{travaglino2026space}. We expect this expression to hold whenever the space evolution is integrable, i.e., after quenches from \emph{integrable initial states}~\cite{Ghoshal:1993tm,PIROLI2017362}, for which the theory in the swapped channel is boundary integrable. 
Considering the additivity originating from  Eq.~\eqref{eq:timeshellSD} it is then possible to sum over time shells to give
\begin{equation}
\label{eq:finalinteracting}
\begin{split}
  \log \operatorname{Tr}[\tilde{\rho}_{\rm st} & \mathrm{e}^{i \sigma \beta_1 \tilde Q_{t_1}} \cdots \mathrm{e}^{i \sigma \beta_n \tilde Q_{t_n}}]  \\
       &= \int_0^t dt'\int \frac{{\rm d}\lambda}{2\pi} \varepsilon'(\mu)\mathcal{L}(\mu,\bar{\beta}(t')).
\end{split}
\end{equation}
Eq.~\eqref{eq:finalinteracting} is a direct generalisation of the expression for the FCS of the charge after quenches from integrable symmetric states proposed in Ref.~\cite{bertini2023nonequilibrium}. The main difference is in the driving term, which now has a time dependence entering through the function $\bar{\beta}(t)$. The presence of interactions, however, preserves the time-shell structure observed in the free case, as the dressing acts separately on each interval $[t_i,t_{i+1}]$. Therefore the $n$-FCS maintains the general form form described in Eq.~\eqref{eq:additivity}, i.e. 
\begin{equation}\label{eq:additivity_interacting}
\begin{split}
    &\mathcal{F}^{(n)}_A(\{t_i,\beta_i\})=i q_0\ell_A \bar{\beta}_0
  +\sum_{j=0}^{n-1} \Delta t_{j} \Gamma(\bar{\beta}_j), \\
  &\Gamma(\bar{\beta})= \int\frac{{\rm d}\mu}{2\pi} \varepsilon'(\mu)
  \left[\mathcal{L}(\mu,\bar{\beta})
  +\mathcal{L}(\mu,-\bar{\beta})\right].
\end{split}
\end{equation}
This closed-form expression can be confirmed numerically considering the paradigmatic example of the XXZ spin-1/2 chain 
\begin{equation}
    H_{\rm XXZ} =  \sum_i \sigma^x_i \sigma^x_{i+1} + \sigma^y_i \sigma^y_{i+1} + \Delta \sigma^z_i \sigma^z_{i+1},
\end{equation} 
where the charge of interest is the magnetisation along the $z$ direction, $Q_A = \sum_{i \in A} \sigma_i^z$.
In this case, to reach system sizes that are large enough to apply space-time duality one has to resort to tensor network based algorithms for the time evolution~\cite{vidal2003simulation,vidal2004,TEBD,cirac2021,ORUS2014117} (see App.~\ref{appC} for more details on our implementation). Since in the XXZ model charge transport is diffusive for $\Delta >1$, we restrict to the gapless regime $\Delta \leq 1$ and, specifically, consider $\Delta = 0.5$ in our numerical evaluations.

\begin{figure}[t]
    \centering
    \includegraphics[width=\linewidth]{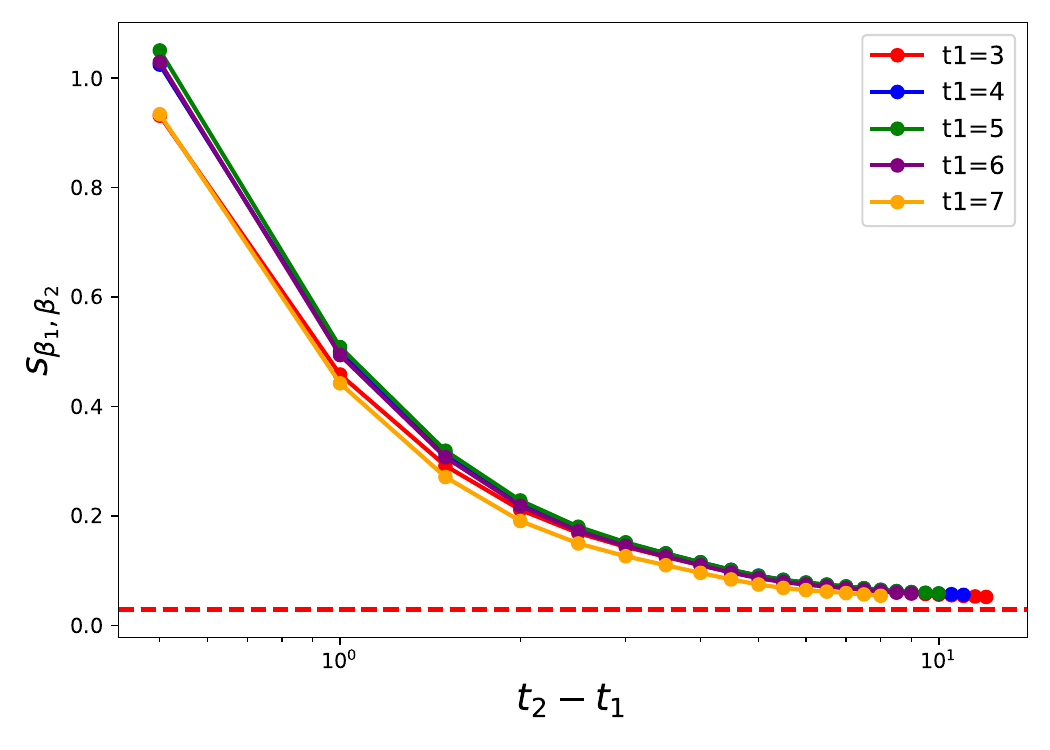}
    \caption{ Slope of the $2-$FCS normalised by the term depending on $t_1$ alone, in the XXZ chain with $\Delta =0.5$. Here, $\beta_1 = \pi/4$ and $\beta_2 = \pi/8$, the total system size is $L=80$ with $\ell_A=40$. The asymptotic value reached by $s_{\beta_1,\beta_2}$, represented by the dashed line is precisely the slope of the single-time FCS obtained in \cite{bertini2023nonequilibrium}.  }
    \label{fig:slope_interacting}
\end{figure}

Focussing on a quench from the N\'eel state $\ket{\uparrow\downarrow\uparrow\downarrow \dots}$, we evaluate the $2$-FCS and check whether it agrees with Eq.~\eqref{eq:additivity_interacting}, i.e., if it takes the factorised form 
\begin{equation}
\!\!\!\mathcal{F}^{(2)}_A(\{t_i,\beta_i\})= t_1\Gamma\left(\beta_1 + \beta_2\right) + (t_2-t_1)\Gamma\left( \beta_{2}\right). 
    \label{eq:additivity-2pt}
\end{equation}
In order to isolate the dependence on the two times, we normalize the FCS by the first term in Eq.~\eqref{eq:additivity-2pt} and consider 
\begin{equation}
 s_{\beta_1,\beta_2} = \frac{1}{\Delta t } \left(\mathcal{F}^{(2)}_A(\{t_i,\beta_i\})-t_1\Gamma\left(\beta_1 + \beta_2\right)\right). 
\end{equation}
The analytical expression in Eq.~\eqref{eq:additivity-2pt} implies that the term in the brackets should (asymptotically) be linear in $\Delta t$ independent of the specific choice of $t_1$, with a coefficient that is the slope of the single time full counting statistics evaluated at $\beta_2$. This is indeed confirmed by the result in Fig.~\ref{fig:slope_interacting}, which is obtained for $\beta_1 = \pi/8$ and $\beta_2=\pi/4$: other examples are discussed in App.~\ref{appC}.

\begin{figure}[t]
    \centering
    \includegraphics[width=\linewidth]{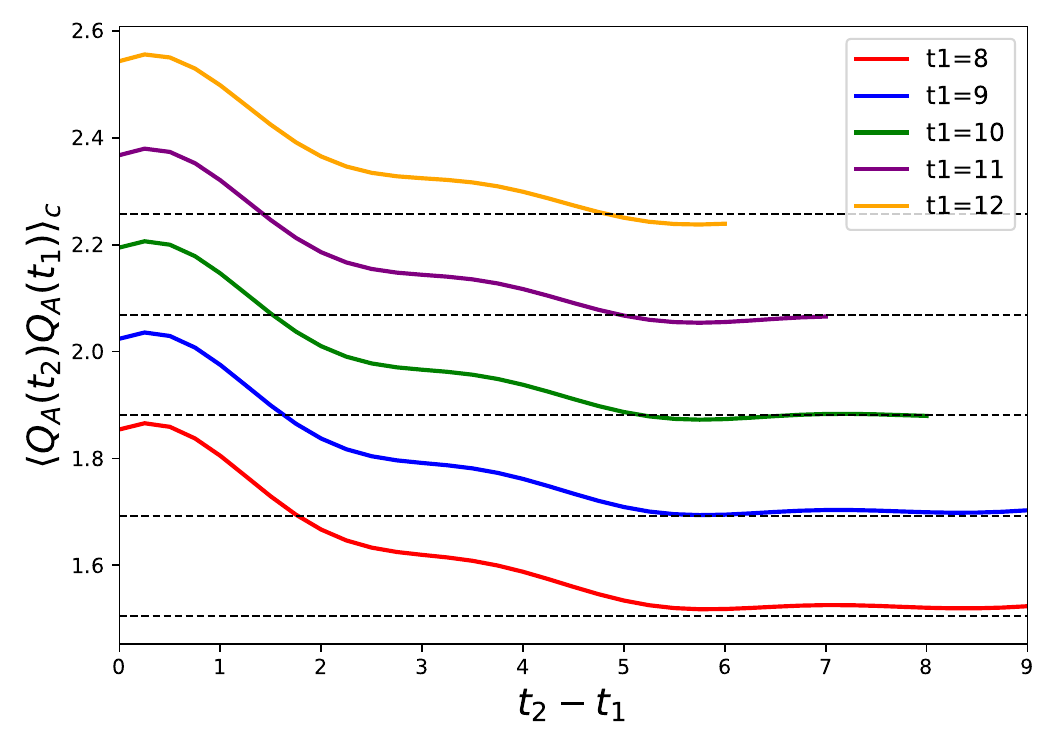}
    \caption{Connected 2-point function for a quench from the Néel state in the XXZ chain with $\Delta = 0.5$, for a subsystem of size $\ell_A=40$. Dashed lines represent the asymptotic prediction $\expval{Q_A^2(t_1)}^{(\infty)}_c$, confirming the prediction for $\Delta t\gtrsim 6$. }
    \label{fig:correlator_interacting}
\end{figure}

Focussing directly on the $2$-FCS allows us to avoid an issue caused by the slow convergence of $\expval*{Q_A(t_1)^2}_c$ to its asymptotic value. Indeed, as shown in App.~\ref{appC}, $\expval*{Q_A(t_1)^2}_c$ approaches its asymptotic value quite slowly and differences of the form 
\be
\expval{Q_A(t_2)Q_A(t_1)}_c -  \expval{Q^2_A(t_1)}_c
\ee
are heavily affected by short time effects. Instead, as we show in Fig.~\ref{fig:correlator_interacting}, the tensor network evolution of $\expval{Q_A(t_2)Q_A(t_1)}_c$ exhibits excellent agreement with the asymptotic prediction
\begin{equation} \label{eq:asymptoticcumulant}
  \expval{Q^2_A(t_1)}_c 
  \approx 2 t_1 \sum_n\int d\mu (q_n^{dr})^2 \rho_n (1-\theta_n ) |v^{dr}_n|, 
\end{equation}
as $t_2-t_1$ increases. Note that this problem did not arise in the free fermionic case, as the approach of the cumulants to their asymptotic value takes place in much faster timescales.

\section{Conclusions}
\label{sec:concl}

In this paper we have characterised the dynamics of time-ordered correlation functions of truncated $U(1)$ charges following quantum quenches. We have shown that in systems with ballistic transport the latter only depend on the first time at which the charges are evaluated, while are completely independent of all the other times. 

This result can be interpreted by saying that the increment of charge between two times $t_1$ and $t_2\geq t_1$ is independent of the charge at $t_1$, that is
\begin{equation}
  \expval{\Delta Q_{A}(t_2,t_1)Q_A(t_1)}=
  \expval{\Delta Q_{A}(t_2,t_1)}\expval{Q_A(t_1)},
\end{equation}
with $\Delta Q_{A}(t_2,t_1)=Q_{A}(t_2)-Q_A(t_1)$.
In this sense it is compatible with the conclusions of Ref.~\cite{Travaglino2025measurements}, where it was shown that, in free fermionic systems, projective measurements of the charge produce a Gaussian distribution of outcomes where the variance is independent of the value of the charge at the previous time. Our present result suggests that this feature should be common to all systems exhibiting ballistic transport. A detailed analysis of projective charge measurements in interacting system, however, is rich in subtleties, and requires much more than the simple probability distribution. This will be addressed in upcoming works.

Another interesting question for future research is to consider charge correlations with different time orderings, for example one could consider the macroscopic OTOCs introduced in Ref.~\cite{Kukuljan} specialised to the case of truncated $U(1)$ charges.

\begin{acknowledgments}
We thank Colin Rylands, Paola Ruggiero and Mario Collura for useful discussions. RT and PC acknowledge support by the ERC-AdG grant MOSE No. 101199196. BB acknowledges support from the Royal Society under the University Research Fellowship No.\ 201101. RT thanks the physics department of the University of Birmingham, where part of this project was carried out.
\end{acknowledgments}

\bibliography{bibliography}

\onecolumngrid
\appendix
\section{Quasiparticle picture for free fermions} \label{appA}
In free fermionic theories of the form $H= \sum_k \varepsilon_k c_k^\dag c_k$, the quasiparticle picture admits an interpretation in terms of ballistic evolution of the density matrix which was developed in Ref.~\cite{Bertini_2018}. The key insight is that the density matrix following a quench from a symmetric integrable state can be expressed approximately as
\begin{equation}
  \rho(0) = \bigotimes_{x,k} \rho_{x,k}
\end{equation}
where $\rho_{x,k}$ is the density matrix of a particle-hole entangled pair, in which the two elements of the pair propagate ballistically in opposite directions,
\begin{eqnarray}
  \label{eq:initialsqueezed}
  \rho_{x,k}=\theta_k\hat{n}_{x,k}(1-\hat{n}_{x,-k})+(1-\theta_k)(1-\hat{n}_{x,k})\hat{n}_{x,-k}
  + \dots~,
\end{eqnarray}
where the dots represent traceless contributions which have no effect on the FCS, and we have coarse grained the fermionic operators within a cell of size $\Delta$,
\begin{equation}
  b_{x,k}^\dag = \sum_{z \in[0,\Delta)}\mathrm{e}^{ikz} c_{x+z}^\dag, \qquad
    \hat n_{x,k} = b_{x,k}^\dag b_{x,k}, 
    \qquad \mathrm{e}^{iHt} b_{x,k}^\dag \mathrm{e}^{-iHt} \approx \mathrm{e}^{-it\varepsilon_k} b_{x + (\partial_k \varepsilon_k) t,k}^\dag.
\end{equation}
The last part of this equation reflects the ballistic evolution of the coarse grained operators, which represents the first order in a $1/\Delta$ expansion \cite{travaglino2026dissipative}, and gives the leading order in the evaluation of all extensive quantities, including the Rényi entropies and the charge fluctuations. 

Combining this semiclassical evolution together with the expression of the
charge in the subsystem 
\begin{equation}
    Q_A = \sum_{x\in A} \sum_z c_{x+z}^\dag c_{x+z} = \sum_{x\in A} \sum_k \hat{n}_{x,k},
\end{equation}
it is possible to evaluate the n-FCS through a computation which mimics the one performed in Ref.~\cite{Travaglino2025measurements}. For each mode the action of the charge is simple, 
\begin{equation}
  \mathrm{e}^{i\beta \hat{n}_{x,k} }\rho_{x,k}(t) = \mathrm{e}^{i\beta} \theta_k\hat{n}_{x,k}(1-\hat{n}_{x,-k})
  +(1-\theta_k)(1-\hat{n}_{x,k})\hat{n}_{x,-k}+\dots.
\end{equation}
Let us now consider the regime in which the times are smaller than $\ell_A/2$ and for simplicity (but without loss of generality) assume that the time-intervals are evenly spaced $t_{j+1}-t_j=\tau$. Then we observe that for each mode $k$, a number $2|v_k|\tau$ of quasiparticles experiences a single charge insertion, the same amount experiences two charge insertion, etc., which leads to
\begin{equation}\label{eq:ffree}
  \mathcal{F}^{(n)}_A(\{t_j,\beta_{j}\}\,) =
  \int \frac{dk}{2\pi} \left[2|v_k|\tau \left(f\left(\sum_{j=1}^n \beta_j,k \right) 
  + f\left(\sum_{j=2}^n \beta,k \right) +\dots \right) \right. +
  \left.  \frac{i}{2} \ell_A \sum_{j=1}^n \beta_j 
  - 2|v_k|\tau \sum_{j=1}^{n} j \beta_j   \right],
\end{equation}
where $t_j = j\tau$, and we have introduced the function
\begin{equation}
  f(\beta,k) = \log \operatorname{Tr}[ \mathrm{e}^{i\beta \hat{n}_{x,k} }\rho_{x,k}]= \log\{1-\theta_k +\theta_k  \mathrm{e}^{i \beta}\} ~.
\end{equation}
Eq.~\eqref{eq:ffree} can be simplified through redefinition of the function to $\tilde f(\beta,k) = \log\{[1-\theta_k]  \mathrm{e}^{i\beta/2}+\theta_k  \mathrm{e}^{i \beta/2}\}$ and observing that for the states \eqref{eq:initialsqueezed} the charge of the initial state is $Q_0=\ell_A/2$, giving
\begin{equation}
 \mathcal{F}^{(n)}_A(\{t_j,\beta_{j}\}\,) =
  i Q_0 \sum_{j=1}^n \beta_j +\int \frac{dk}{2\pi} 
  \left\{2|v_k|\tau \left(\tilde f\left(\sum_{j=1}^n \beta_j,k \right) 
  + \tilde f\left(\sum_{j=2}^n \beta,k \right) +\dots \right)\right\}\, .
\end{equation}
It is then a matter of straightforward algebra to convince oneself that this is equivalent to~\eqref{eq:Fdimerfinal}
\begin{equation}
\mathcal{F}^{(n)}_A(\{t_j,\beta_j\})= i Q_0 \bar\beta(0)
  +\sum_{\sigma=\pm} \int\frac{{\rm d}k}{2\pi}   v_k 
  \int_0^t dt'\mathcal{L}\left(k,\sigma\bar{\beta}(t')\right).
\end{equation}
The last expression can be easily generalised to arbitrary time values
$(t_1,\dots t_n)$ by using the definition of the function $\bar{\beta}(t')$ in
Eq.~\eqref{eq:defBarBeta}.

\section{Rule 54}\label{sec:R54App}
\subsection{Setting}
We consider a qubit chain of size $L$ (even) with periodic boundary conditions. The dynamics is defined in discrete time-steps, so that at odd and even time-steps the three-site gates centred around sites with opposite parity
\begin{equation}
  \ket{\psi_{t}}=\mathbb{U}_t\ket{\psi_{t-1}},\qquad
  \mathbb{U}_t=\begin{cases}
    U_{1,2,3}U_{3,4,5}\cdots U_{L-1,L,1},\quad& t\equiv 0\pmod 2,\\
    U_{2,3,4}U_{4,5,6}\cdots U_{L,1,2},&t\equiv 1\pmod 2,
  \end{cases}
\end{equation}
where $U_{j,j+1,j+2}$ is a three-site unitary update with support on sites $j$, $j+1$, and $j+2$
\begin{equation}
\begin{gathered}
  U_{j,j+1,j+2}=\1^{\otimes j-1} \otimes U \otimes \1^{\otimes L-j-2},\qquad
  \mel{b_1 b_2 b_3}{U}{s_1 s_2 s_3}=\delta_{s_1,b_1}\delta_{s_3,b_3}
  \delta_{b_2,\chi(s_1,s_2,s_3)},\\
  \chi(s_1,s_2,s_3)\equiv s_1+s_2+s_3+s_1 s_3\pmod{2},
\end{gathered}
\end{equation}
and $\chi:\mathbb{Z}_2^{\times 3}\to \mathbb{Z}_2$ is the three-site deterministic update rule.

The extensive charge we will use here is $Q^{-}$, which corresponds to the imbalance between the number of left and right moving particles
\begin{equation}
  Q^{-}=\sum_{j=1}^{L}(-1)^j q^{-}_{j,j+1},\qquad
  q^{-}_{j,j+1}=\frac{1}{4}\left(1-\sigma_{j}^z\right)\left(1-\sigma_{j+1}^{z}\right).
\end{equation}
Note that for the solvable quench we are considering here, the initial state (to be specified below) is \emph{symmetric} under the charge $Q^{-}$.

To express the $n$-FCS we define the trace of the reduced density matrix obtained by applying $\mathrm{e}^{\alpha Q^{-}_A}$ between every time-step of the evolution 
\begin{equation}
  Z_{A}(\underline{\alpha})=
  \Tr\left[
    \mathrm{e}^{-(-1)^{t}\alpha_t Q^{-}_{A}}
  \mathbb{U}_t
  \mathrm{e}^{-(-1)^{t-1}\alpha_{t-1} Q^{-}_{A}}
  \mathbb{U}_{t-1}
  \cdots
  \mathrm{e}^{\alpha_1 Q^{-}_{A}}
  \mathbb{U}_1
  \ketbra{\Psi_0}{\Psi_0}
  \mathbb{U}_1^{\dagger}
  \mathbb{U}_2^{\dagger}
  \cdots
  \mathbb{U}_t^{\dagger}\right].
\end{equation}
Since $Q^{-}$ commutes with the product $\mathbb{U}_2\mathbb{U}_1$, but anticommutes with a single-step time-evolution operator, here we have conveniently changed the sign of $Q_A$ after each time-step. This object is a slight generalisation of what is needed for $n$-FCS, which is obtained from $Z_{A}(\underline{\alpha})$ by appropriately choosing the $t$-tuple $\underline{\alpha}=(\alpha_1,\alpha_2,\ldots,\alpha_t)$
\begin{equation}\label{eq:asymptFAn}
  \mathcal{F}_{A}^{(n)}(\{t_i,\beta_i\})=
  \log Z_A\big(\underbrace{0,\ldots,0,i(-1)^{t_1}\beta_1}_{t_1},
    \underbrace{0,\ldots,0,i(-1)^{t_2}\beta_2}_{t_2-t_1},
    \underbrace{0,\ldots,0,i(-1)^{t_3}\beta_3}_{t_3-t_2},\ldots\big).
\end{equation}

To be able to graphically express $Z_{A}(\underline{\alpha})$ we introduce the following two tensors
\begin{equation}
 \begin{tikzpicture}[baseline={([yshift=-0.6ex]current bounding box.center)},scale=0.5]
      \node at (-1.275,0) {{$s_1$}};
      \node at (0,1.275) {{$s_2$}};
      \node at (1.275,0) {{$s_3$}};
      \node at (0,-1.275) {{$s_4$}};
      \gridLine{-1}{0}{1}{0}
      \gridLine{0}{-1}{0}{1}
      \bdCircle{0}{0}{colU}
    \end{tikzpicture}=\delta_{s_2,\chi(s_1,s_4,s_3)},\qquad
 \begin{tikzpicture}[baseline={([yshift=-0.6ex]current bounding box.center)},scale=0.5]
      \node at (-1.275,0) {{$s_1$}};
      \node at (0,1.275) {{$s_2$}};
      \node at (1.275,0) {{$s_3$}};
      \node at (0,-1.275) {{$s_4$}};
      \gridLine{-1}{0}{1}{0}
      \gridLine{0}{-1}{0}{1}
      \bdCircle{0}{0}{colU}
    \end{tikzpicture}=\delta_{s_2,\chi(s_1,s_4,s_3)}
    \left[1+\delta_{s_1+s_2,2} (\mathrm{e}^{-\alpha}-1)\right]
    \left[1+\delta_{s_2+s_3,2} (\mathrm{e}^{\alpha}-1)\right],
\end{equation}
in terms of which we can express the local time-evolution operator as
\begin{equation}
  U=
    \begin{tikzpicture}[baseline={([yshift=-0.6ex]current bounding box.center)},scale=0.5]
      \gridLine{-1}{0}{1}{0}
      \gridLine{-1}{-1}{-1}{1}
      \gridLine{0}{-1}{0}{1}
      \gridLine{1}{-1}{1}{1}
      \bCircle{0}{0}{colU}
    \end{tikzpicture}\,,\qquad
    \mathrm{e}^{\alpha (-q_{1,2}+ q_{2,3})}
    U=
    \begin{tikzpicture}[baseline={([yshift=-0.6ex]current bounding box.center)},scale=0.5]
      \gridLine{-1}{0}{1}{0}
      \gridLine{-1}{-1}{-1}{1}
      \gridLine{0}{-1}{0}{1}
      \gridLine{1}{-1}{1}{1}
      \bdCircle{0}{0}{colU}
    \end{tikzpicture}\,.
\end{equation}
We also note that we need two additional tensors to account for different parities of the first and last site of $A$
\begin{equation}
 \begin{tikzpicture}[baseline={([yshift=-0.6ex]current bounding box.center)},scale=0.5]
      \node at (-1.275,0) {{$s_1$}};
      \node at (0,1.275) {{$s_2$}};
      \node at (1.275,0) {{$s_3$}};
      \node at (0,-1.275) {{$s_4$}};
      \gridLine{-1}{0}{1}{0}
      \gridLine{0}{-1}{0}{1}
      \bdLCircle{0}{0}{colU}
    \end{tikzpicture}=\delta_{s_2,\chi(s_1,s_4,s_3)}
    \left[1+\delta_{s_2+s_3,2} (\mathrm{e}^{\alpha}-1)\right],\qquad
 \begin{tikzpicture}[baseline={([yshift=-0.6ex]current bounding box.center)},scale=0.5]
      \node at (-1.275,0) {{$s_1$}};
      \node at (0,1.275) {{$s_2$}};
      \node at (1.275,0) {{$s_3$}};
      \node at (0,-1.275) {{$s_4$}};
      \gridLine{-1}{0}{1}{0}
      \gridLine{0}{-1}{0}{1}
      \bdRCircle{0}{0}{colU}
    \end{tikzpicture}=\delta_{s_2,\chi(s_1,s_4,s_3)}
    \left[1+\delta_{s_1+s_2,2} (\mathrm{e}^{-\alpha}-1)\right],
\end{equation}
which gives us the following expression for the tensors at the boundaries
\begin{equation}
    \mathrm{e}^{-\alpha q_{1,2}} U=
    \begin{tikzpicture}[baseline={([yshift=-0.6ex]current bounding box.center)},scale=0.5]
      \gridLine{-1}{0}{1}{0}
      \gridLine{-1}{-1}{-1}{1}
      \gridLine{0}{-1}{0}{1}
      \gridLine{1}{-1}{1}{1}
      \bdLCircle{0}{0}{colU}
    \end{tikzpicture}
    \,,\qquad
    \mathrm{e}^{\alpha q_{2,3}} U=
    \begin{tikzpicture}[baseline={([yshift=-0.6ex]current bounding box.center)},scale=0.5]
      \gridLine{-1}{0}{1}{0}
      \gridLine{-1}{-1}{-1}{1}
      \gridLine{0}{-1}{0}{1}
      \gridLine{1}{-1}{1}{1}
      \bdRCircle{0}{0}{colU}
    \end{tikzpicture}.
\end{equation}
Using this notation, the full time-evolution step is 
\begin{equation}
  \mathrm{e}^{-\alpha_2 Q^{-}_A} \mathbb{U}_2\mathrm{e}^{\alpha_1 Q^{-}_A} \mathbb{U}_1=
  \begin{tikzpicture}[baseline={([yshift=-0.6ex]current bounding box.center)},scale=0.5]
    \def\X{12}
    \foreach\x in {1,...,\X}{\gridLine{\x}{0}{\x}{3}}
    \gridLine{0.35}{1}{\X+0.65}{1}
    \gridLine{0.35}{2}{\X+0.65}{2}
    \foreach\x in {1,3,...,\X}{
    }
    \foreach\x in {1,3,...,5}{
      \bdCircle{\x}{2}{colU}
      \bdCircle{\x+1}{1}{colU}
    }
    \foreach \x in {7,9,...,\X}{
      \bCircle{\x}{2}{colU}
      \bCircle{\x+1}{1}{colU}
    }
    \leftHook{0.35}{2}
    \leftHook{0.35}{1}
    \rightHook{\X+0.65}{2}
    \rightHook{\X+0.65}{1}
    \bdLCircle{1}{2}{colU}
    \bdRCircle{6}{1}{colU}
    \node[anchor=west] at ({\X+0.75},2) {$2$};
    \node[anchor=west] at ({\X+0.75},1) {$1$};
  \end{tikzpicture},
\end{equation}
where the label on the right denotes the subscript $j$ of the $\alpha_j$ used, and is assumed to be the same in the full row. In this example the subsystem $A$ is given by the left half of the chain, and we assume periodic boundary conditions. Additionally we denote the transpose with a different colour
\begin{equation}
  U^{\dagger}=
    \begin{tikzpicture}[baseline={([yshift=-0.6ex]current bounding box.center)},scale=0.5]
      \gridLine{-1}{0}{1}{0}
      \gridLine{-1}{-1}{-1}{1}
      \gridLine{0}{-1}{0}{1}
      \gridLine{1}{-1}{1}{1}
      \bCircle{0}{0}{colUc}
    \end{tikzpicture}\,.
\end{equation}

Using these graphical conventions, we can express $Z_{A}(\underline{\alpha})$ as
\begin{equation}\label{eq:densityMatrixTN}
    Z_{A}(\underline{\alpha})=
  \begin{tikzpicture}[baseline={([yshift=-0.6ex]current bounding box.center)},scale=0.5]
    \def\X{12}
    \def\Y{6}
    \foreach\x in {1,...,\X}{\gridLine{\x}{0}{\x}{\Y+0.65}}
    \foreach\x in {1,...,\X}{\gridLine{\x}{-0.75}{\x}{-\Y-0.65-0.75}}
    \foreach \x in {1,3,...,\X}{
      \vmpsV{\x}{0}{colIst}{colLines}
      \vmpsW{\x+1}{0}{colIst}{colLines}
      \vmpsV{\x}{-0.75}{colIstC}{colLines}
      \vmpsW{\x+1}{-0.75}{colIstC}{colLines}
    }
    \foreach \x in {1,3,...,\X}
    {
      \topHook{\x}{\Y+0.65}
      \topHook{\x+1}{\Y+0.65}
      \botHook{\x}{-\Y-1.4}
      \botHook{\x+1}{-\Y-1.4}
    }
    \foreach \t in {1,3,...,\Y}
    {
      \gridLine{0.35}{\t}{\X+0.65}{\t}
      \gridLine{0.35}{1+\t}{\X+0.65}{1+\t}
      \gridLine{0.35}{-\t-0.75}{\X+0.65}{-\t-0.75}
      \gridLine{0.35}{-1.75-\t}{\X+0.65}{-1.75-\t}
      \foreach\x in {1,3,...,5}{
        \bdCircle{\x}{1+\t}{colU}
        \bdCircle{\x+1}{\t}{colU}
        \bCircle{\x}{-1.75-\t}{colUc}
        \bCircle{\x+1}{-0.75-\t}{colUc}
      }
      \foreach \x in {7,9,...,\X}{
        \bCircle{\x}{1+\t}{colU}
        \bCircle{\x+1}{\t}{colU}
        \bCircle{\x}{-1.75-\t}{colUc}
        \bCircle{\x+1}{-0.75-\t}{colUc}
      }
      \leftHook{0.35}{1+\t}
      \leftHook{0.35}{\t}
      \leftHook{0.35}{-1.75-\t}
      \leftHook{0.35}{-0.75-\t}
      \rightHook{\X+0.65}{1+\t}
      \rightHook{\X+0.65}{\t}
      \rightHook{\X+0.65}{-1.75-\t}
      \rightHook{\X+0.65}{-0.75-\t}
      \bdLCircle{1}{1+\t}{colU}
      \bdRCircle{6}{\t}{colU}
      \bCircle{1}{-1.75-\t}{colUc}
      \bCircle{6}{-0.75-\t}{colUc}
    }
    \node[anchor=west] at ({\X+0.75},\Y) {$t$};
    \node[anchor=west] at ({\X+0.75},\Y-1) {$\vdots$};
    \node[anchor=west] at ({\X+0.75},3) {$\vdots$};
    \node[anchor=west] at ({\X+0.75},2) {$2$};
    \node[anchor=west] at ({\X+0.75},1) {$1$};
  \end{tikzpicture}\xrightarrow{|\bar{A}|\to\infty}
\begin{tikzpicture}[baseline={([yshift=-0.6ex]current bounding box.center)},scale=0.5]
    \def\X{12}
    \def\Y{6}
    \foreach \t in {1,...,\Y}
    {
      \draw [thick,colLines,rounded corners=2] ({-0.5-\t*0.2},{\t}) rectangle ({7.5+0.2*\t},{-0.75-\t});
    }
    \foreach\x in {1,...,6}{\gridLine{\x}{0}{\x}{\Y+0.65}}
    \foreach\x in {1,...,6}{\gridLine{\x}{-0.75}{\x}{-\Y-0.65-0.75}}
    \foreach \x in {1,3,...,5}{
      \vmpsV{\x}{0}{colIst}{colLines}
      \vmpsW{\x+1}{0}{colIst}{colLines}
      \vmpsV{\x}{-0.75}{colIstC}{colLines}
      \vmpsW{\x+1}{-0.75}{colIstC}{colLines}
    }
    \mpsWire{0}{0}{0}{6.75}
    \mpsWire{7}{0}{7}{6.75}
    \mpsBvecW{0}{0}{colMPS}
    \mpsBvecV{7}{0}{colMPS}
    \mpsBvec{0}{6.75}
    \mpsBvec{7}{6.75}
    \foreach \t in {1,3,...,\Y}
    {
      \foreach\x in {1,3,...,5}{
        \bdCircle{\x}{1+\t}{colU}
        \bdCircle{\x+1}{\t}{colU}
        \bCircle{\x}{-1.75-\t}{colUc}
        \bCircle{\x+1}{-0.75-\t}{colUc}
      }
      \bdLCircle{1}{1+\t}{colU}
      \bdRCircle{6}{\t}{colU}
      \bCircle{1}{-1.75-\t}{colUc}
      \bCircle{6}{-0.75-\t}{colUc}
      \mpsA{0}{\t}{colMPS}
      \mpsB{0}{\t+1}{colMPS}
      \mpsB{7}{\t}{colMPS}
      \mpsA{7}{\t+1}{colMPS}
    }
    \foreach \x in {1,3,5}
    {
      \topHook{\x}{\Y+0.65}
      \topHook{\x+1}{\Y+0.65}
      \botHook{\x}{-\Y-1.4}
      \botHook{\x+1}{-\Y-1.4}
    }
    \node[anchor=west] at ({8.75},\Y) {$t$};
    \node[anchor=west] at ({8.75},\Y-1) {$\vdots$};
    \node[anchor=west] at ({8.75},3) {$\vdots$};
    \node[anchor=west] at ({8.75},2) {$2$};
    \node[anchor=west] at ({8.75},1) {$1$};
  \end{tikzpicture},
\end{equation}
where the initial state is taken to be a \emph{solvable} product state~\cite{klobas2021exactrelaxation}
\begin{equation}\label{eq:defSolIS}
  \begin{tikzpicture}[baseline={([yshift=-0.6ex]current bounding box.center)},scale=0.5]
    \gridLine{0}{0}{0}{0.75}
    \vmpsV{0}{0}{colIst}{colLines}
    \end{tikzpicture}=\begin{bmatrix}1\\0\end{bmatrix},
  \qquad
  \begin{tikzpicture}[baseline={([yshift=-0.6ex]current bounding box.center)},scale=0.5]
    \gridLine{0}{0}{0}{0.75}
    \vmpsW{0}{0}{colIst}{colLines}
    \end{tikzpicture}=\begin{bmatrix}\sqrt{1-\vartheta}\\\sqrt{\vartheta}\end{bmatrix},\quad
      0<\vartheta<1.
\end{equation}
    The r.h.s.\ above follows from the fact that whenever the rest of the system is large enough compared to time $t$, the action of the rest of the system on our subsystem of choice can be encoded in the fixed points of the space transfer-matrix~\cite{klobas2021exact,lerose2021influence,bertini_negativity}, and for the solvable initial state in Eq.~\eqref{eq:defSolIS} the fixed points take a matrix-product form with bond dimension $3$~\cite{klobas2021exact,klobas2021exactrelaxation}, with the precise forms of the tensors reported in App.~\ref{sec:FPtensors}.

\subsection{Asymptotic $n$-FCS for large subsystems}
\subsubsection{Fixed points}

Whenever we are interested in timescales that are short compared to the subsystem size $|A|$, we are able to use a simplification similar to the one leading to the r.h.s.\ of Eq.~\eqref{eq:densityMatrixTN} (see App.~\ref{sec:FPtensors}) so that $Z_{A}(\underline{\alpha})$ factorises into two contributions coming from the edges of the subsystem
\begin{equation}\label{eq:R54ZwithFPs}
  \left.Z_A(\underline{\alpha})\right|_{2t<|A|} = 
  z_{-}(\underline{\alpha}) z_{+}(\underline{\alpha}),\qquad
  z_{-}(\underline{\alpha})=
  \begin{tikzpicture}[baseline={([yshift=-0.6ex]current bounding box.center)},scale=0.5]
    \def\X{1}
    \def\Y{6}
    \foreach \t in {1,...,\Y}
    {
      \draw [thick,colLines,rounded corners=2] ({-0.5-\t*0.2},{\t}) rectangle ({1.5+0.2*\t},{-0.25-0.2*\t});
    }
    \node[anchor=west] at (2.75,1) {$1$};
    \node[anchor=west] at (2.75,2) {$2$};
    \node[anchor=west] at (2.75,3) {$3$};
    \node[anchor=west] at (2.75,4.5) {$\vdots$};
    \node[anchor=west] at (2.75,6) {$t$};
    \mpsWire{0}{0}{0}{6.75}
    \mpsWire{1}{0}{1}{6.75}
    \mpsBvecW{0}{0}{colMPS}
    \mpsBvecV{1}{0}{colMPSBetaB}
    \mpsBvec{0}{6.75}
    \mpsBvec{1}{6.75}
    \foreach \t in {1,3,...,\Y}
    {
      \mpsA{0}{\t}{colMPS}
      \mpsB{0}{\t+1}{colMPS}
      \mpsB{1}{\t}{colMPSBetaB}
      \mpsA{1}{\t+1}{colMPSBetaB}
    }
    \foreach \t in {0,...,\Y}{\wedgeR{1}{\t}}
  \end{tikzpicture}\,,\quad
  z_{+}(\underline{\alpha})=
  \begin{tikzpicture}[baseline={([yshift=-0.6ex]current bounding box.center)},scale=0.5]
    \def\X{1}
    \def\Y{6}
    \foreach \t in {1,...,\Y}
    {
      \draw [thick,colLines,rounded corners=2] ({-0.5-\t*0.2},{\t}) rectangle ({1.5+0.2*\t},{-0.25-0.2*\t});
    }
    \mpsWire{0}{0}{0}{6.75}
    \mpsWire{1}{0}{1}{6.75}
    \mpsBvecW{0}{0}{colMPSBetaB}
    \mpsBvecV{1}{0}{colMPS}
    \mpsBvec{0}{6.75}
    \mpsBvec{1}{6.75}
    \foreach \t in {1,3,...,\Y}
    {
      \mpsB{1}{\t}{colMPS}
      \mpsA{1}{\t+1}{colMPS}
      \mpsA{0}{\t}{colMPSBetaB}
      \mpsB{0}{\t+1}{colMPSBetaB}
    }
    \node[anchor=west] at (2.75,1) {$1$};
    \node[anchor=west] at (2.75,2) {$2$};
    \node[anchor=west] at (2.75,3) {$3$};
    \node[anchor=west] at (2.75,4.5) {$\vdots$};
    \node[anchor=west] at (2.75,6) {$t$};
    \foreach \t in {0,...,\Y}{\wedgeL{0}{\t}}
  \end{tikzpicture}\,.
\end{equation}
The light-blue fixed-point tensors inherit their dependence on the position on the time-lattice from the bulk tensors, and at the step $j$ they are given by
\begin{equation}
\begin{aligned}
  \begin{tikzpicture}[baseline={([yshift=-0.6ex]current bounding box.center)},scale=0.5]
    \mpsWire{0}{-0.75}{0}{0.75}
    \gridLine{-0.75}{0}{0.75}{0}
    \mpsA{0}{0}{colMPSBetaB}
    \wedgeL{0}{0}
    \node at (-1,0) {$0$};
    \node at (1,0) {$0$};
  \end{tikzpicture}&=
  \begin{bmatrix}
    1-\vartheta & 1-\vartheta & -(1-\vartheta) \\
    \Pi_j \vartheta & \Pi_j \vartheta & \Pi_j (1-\vartheta) \\
    \vartheta & - \frac{\vartheta^2}{1-\vartheta} & - \vartheta
  \end{bmatrix},&\qquad
  \begin{tikzpicture}[baseline={([yshift=-0.6ex]current bounding box.center)},scale=0.5]
    \mpsWire{0}{-0.75}{0}{0.75}
    \gridLine{-0.75}{0}{0.75}{0}
    \mpsA{0}{0}{colMPSBetaB}
    \wedgeL{0}{0}
    \node at (-1,0) {$1$};
    \node at (1,0) {$1$};
  \end{tikzpicture}&=
  \begin{bmatrix}
    0 & 1 & 0 \\
    \Pi_{j} & 0 & 0 \\
    0 & 0 & 0
  \end{bmatrix},\\
  \begin{tikzpicture}[baseline={([yshift=-0.6ex]current bounding box.center)},scale=0.5]
    \mpsWire{0}{-0.75}{0}{0.75}
    \gridLine{-0.75}{0}{0.75}{0}
    \mpsA{0}{0}{colMPSBetaB}
    \wedgeL{0}{0}
    \node at (-1,0) {$0$};
    \node at (1,0) {$1$};
  \end{tikzpicture}&=
  \begin{tikzpicture}[baseline={([yshift=-0.6ex]current bounding box.center)},scale=0.5]
    \mpsWire{0}{-0.75}{0}{0.75}
    \gridLine{-0.75}{0}{0.75}{0}
    \mpsA{0}{0}{colMPSBetaB}
    \wedgeL{0}{0}
    \node at (-1,0) {$1$};
    \node at (1,0) {$0$};
  \end{tikzpicture}=
  \begin{bmatrix}
    0 & 1-\vartheta & - (1-\vartheta) \\
    \Pi_j \vartheta& 0 & 0 \\
    \vartheta & 0 & 0 
  \end{bmatrix},&
  \begin{tikzpicture}[baseline={([yshift=-0.6ex]current bounding box.center)},scale=0.5]
    \mpsWire{0}{-0.75}{0}{0.75}
    \gridLine{-0.75}{0}{0.75}{0}
    \mpsB{0}{0}{colMPSBetaB}
    \wedgeL{0}{0}
    \node at (-1,0) {$s$};
    \node at (1,0) {$b$};
  \end{tikzpicture}&=
  \begin{bmatrix}
    \delta_{s,0} & 0 & 0 \\ 0 & \delta_{s,1} & 0 \\ 0 & 0 & \delta_{s,1} \Pi_{j-1}
  \end{bmatrix},\\
  \begin{tikzpicture}[baseline={([yshift=-0.6ex]current bounding box.center)},scale=0.5]
    \mpsWire{0}{0}{0}{0.75}
    \mpsBvecW{0}{0}{colMPSBetaB}
    \wedgeL{0}{0}
  \end{tikzpicture}&=
  \begin{bmatrix} 1 \\ 0 \\ 0 \end{bmatrix},\quad
  \begin{tikzpicture}[baseline={([yshift=-0.6ex]current bounding box.center)},scale=0.5]
    \mpsWire{0}{0}{0}{0.75}
    \mpsBvecV{0}{0}{colMPSBetaB}
    \wedgeL{0}{0}
  \end{tikzpicture}=
  \begin{bmatrix} 1-\vartheta \\ \Pi_0 \vartheta \\ -\frac{\vartheta^2}{1-\vartheta}\end{bmatrix},\quad
  \begin{tikzpicture}[baseline={([yshift=-0.6ex]current bounding box.center)},scale=0.5]
    \mpsWire{0}{-0.75}{0}{0}
    \mpsBvec{0}{0}
    \end{tikzpicture}=\begin{bmatrix} 1 & 1 & 0 \end{bmatrix},&\qquad
    \Pi_j&=\prod_{l=j+1}^t \mathrm{e}^{\alpha_l},
\end{aligned}
\end{equation}
and the different sides are related to each other through the mapping $\Pi_j\leftrightarrow \Pi_j^{-1}$
\begin{equation}
  \left\{
  \begin{tikzpicture}[baseline={([yshift=-0.6ex]current bounding box.center)},scale=0.5]
    \mpsWire{0}{-0.75}{0}{0.75}
    \gridLine{-0.75}{0}{0.75}{0}
    \mpsA{0}{0}{colMPSBetaB}
    \wedgeL{0}{0}
  \end{tikzpicture},
  \begin{tikzpicture}[baseline={([yshift=-0.6ex]current bounding box.center)},scale=0.5]
    \mpsWire{0}{-0.75}{0}{0.75}
    \gridLine{-0.75}{0}{0.75}{0}
    \mpsB{0}{0}{colMPSBetaB}
    \wedgeL{0}{0}
  \end{tikzpicture},
  \begin{tikzpicture}[baseline={([yshift=-0.6ex]current bounding box.center)},scale=0.5]
    \mpsWire{0}{0}{0}{0.75}
    \mpsBvecW{0}{0}{colMPSBetaB}
    \wedgeL{0}{0}
  \end{tikzpicture},
  \begin{tikzpicture}[baseline={([yshift=-0.6ex]current bounding box.center)},scale=0.5]
    \mpsWire{0}{0}{0}{0.75}
    \mpsBvecV{0}{0}{colMPSBetaB}
    \wedgeL{0}{0}
  \end{tikzpicture}\right\}
  \xleftrightarrow{\Pi_j\leftrightarrow \frac{1}{\Pi_j}}
  \left\{
  \begin{tikzpicture}[baseline={([yshift=-0.6ex]current bounding box.center)},scale=0.5]
    \mpsWire{0}{-0.75}{0}{0.75}
    \gridLine{-0.75}{0}{0.75}{0}
    \mpsA{0}{0}{colMPSBetaB}
    \wedgeR{0}{0}
  \end{tikzpicture},
  \begin{tikzpicture}[baseline={([yshift=-0.6ex]current bounding box.center)},scale=0.5]
    \mpsWire{0}{-0.75}{0}{0.75}
    \gridLine{-0.75}{0}{0.75}{0}
    \mpsB{0}{0}{colMPSBetaB}
    \wedgeR{0}{0}
  \end{tikzpicture},
  \begin{tikzpicture}[baseline={([yshift=-0.6ex]current bounding box.center)},scale=0.5]
    \mpsWire{0}{0}{0}{0.75}
    \mpsBvecW{0}{0}{colMPSBetaB}
    \wedgeR{0}{0}
  \end{tikzpicture},
  \begin{tikzpicture}[baseline={([yshift=-0.6ex]current bounding box.center)},scale=0.5]
    \mpsWire{0}{0}{0}{0.75}
    \mpsBvecV{0}{0}{colMPSBetaB}
    \wedgeR{0}{0}
  \end{tikzpicture}\right\}
\xrightarrow{\Pi_j\to 1}=
  \left\{
  \begin{tikzpicture}[baseline={([yshift=-0.6ex]current bounding box.center)},scale=0.5]
    \mpsWire{0}{-0.75}{0}{0.75}
    \gridLine{-0.75}{0}{0.75}{0}
    \mpsA{0}{0}{colMPS}
  \end{tikzpicture},
  \begin{tikzpicture}[baseline={([yshift=-0.6ex]current bounding box.center)},scale=0.5]
    \mpsWire{0}{-0.75}{0}{0.75}
    \gridLine{-0.75}{0}{0.75}{0}
    \mpsB{0}{0}{colMPS}
  \end{tikzpicture},
  \begin{tikzpicture}[baseline={([yshift=-0.6ex]current bounding box.center)},scale=0.5]
    \mpsWire{0}{0}{0}{0.75}
    \mpsBvecW{0}{0}{colMPS}
  \end{tikzpicture},
  \begin{tikzpicture}[baseline={([yshift=-0.6ex]current bounding box.center)},scale=0.5]
    \mpsWire{0}{0}{0}{0.75}
    \mpsBvecV{0}{0}{colMPS}
  \end{tikzpicture}\right\}.
\end{equation}
Note that at position $j$ the fixed point tensors depend \emph{only} on the sum of $\alpha_{l}$ with $l>j$. The algebraic relations leading to this form of fixed points is given in App.~\ref{sec:FPtensors}.

\subsubsection{$n$-FCS for sparse measurements}
We can equivalently rewrite the above contributions to the $n$-FCS as the following matrix products
\begin{equation}
  z_{+}(\underline{\alpha})=
  \mel{T}{M_{+}(\Pi_{2\lfloor{\frac{t+1}{2}}\rfloor-1})
  \cdots M_{+}(\Pi_3)M_{+}(\Pi_1)}{B_{+}},\quad
  z_{-}(\underline{\alpha})=
  \mel{T}{M_{-}(\Pi_{2\lfloor{\frac{t}{2}}\rfloor})\cdots M_{-}(\Pi_2)M_{-}(\Pi_0)}{B_{-}},
\end{equation}
where the relevant objects can be reduced to be $3$-dimensional (from $9$-dimensional)
\begin{equation}
  M_{+}(\Pi)=
    \begin{bmatrix}
      (1-\vartheta)^2&\Pi&1-\vartheta\\
      \vartheta(1-\vartheta)&0&\vartheta\\
      \Pi\vartheta&0&0
    \end{bmatrix},\qquad
    M_{-}(\Pi)=
    \begin{bmatrix}
      (1-\vartheta)^2&1&1-\vartheta\\
      \Pi^{-1}\vartheta(1-\vartheta)&0&\Pi^{-1}\vartheta\\
      \Pi^{-1}\vartheta&0&0
    \end{bmatrix},
\end{equation}
and the boundary vectors are
\begin{equation}
  \bra{T}=\begin{bmatrix} 1 & 1 & 1 \end{bmatrix},\qquad
    \ket{B_{+}}=\begin{bmatrix} 1-\vartheta\\\vartheta\\0\end{bmatrix},\qquad
      \ket{B_{-}}=\begin{bmatrix} 0\\0\\1\end{bmatrix}.
\end{equation}
Note that we have for convenience artificially added an additional matrix $M_{+}(\Pi_0)$ by appropriately defining the state $\ket{B_{+}}$ (explicitly, we multiplied the ``naive'' bottom boundary vector by the inverse of $M_{+}(\Pi_0)$).

These expressions further simplify in the sparse limit of Eq.~\eqref{eq:asymptFAn}, in which case we specialise $\Pi_j$ to
\begin{equation}
  \Pi_{t_{n}}=1,\qquad 
  \Pi_{t_{j}-1}=\Pi_{t_{j}-2}=\ldots=\Pi_{t_{j-1}}=
  \mathrm{e}^{i\sum_{l=j}^n (-1)^{t_l} \beta_{l}}.
\end{equation}
Now let us also assume that all $t_j$ are even, as $Q^{(-)}$ is only conserved after $2$ time-steps. Then we get
\begin{equation}
    z_{\pm} =
  \mel{T}{
    \left(M_{\pm}(\Pi_{t_{n-1}})\right)^{\frac{t_n-t_{n-1}}{2}}
    \cdots
    \left(M_{\pm}(\Pi_{t_1})\right)^{\frac{t_2-t_1}{2}}
    \left(M_{\pm}(\Pi_0)\right)^{\frac{t_1}{2}}
    }{B_{\pm}}.
\end{equation}
So far the result is exact and holds for any $\{t_j\}_j$, but in the limit of large $t_{j+1}-t_j$ we can approximate the above products with the contribution coming from the leading eigenvalues of each matrix. Simplifying the notation by making the change $\Pi_{t_j}\to \Pi_j$ we get the expressions reported in the main text.

\subsection{Local algebraic relations and fixed-point tensors}\label{sec:FPtensors}
Here we report the details of simplification leading to Eq.~\eqref{eq:R54ZwithFPs} (and, as a special case, also to the r.h.s.\ of Eq.~\eqref{eq:densityMatrixTN}). We start by noting that we need to introduce two sets of fixed-point tensors, which encode the effect of decorated gates away from the boundary of the subsystem (i.e.\ those with the circular mark in Eq.~\eqref{eq:densityMatrixTN}). We denote them with the orange colour and we impose that they satisfy the following set of local algebraic relations
\begin{equation}
  \begin{gathered}
  \begin{tikzpicture}[baseline={([yshift=-0.6ex]current bounding box.center)},scale=0.5]
    \mpsWire{0}{0.25}{0}{4.75}
    \foreach \t in {1,...,4}{
      \draw [thick,colLines,rounded corners=2] ({1.75},{-0.75-\t}) -- ({-0.5-\t*0.2},{-0.75-\t}) -- ({-0.5-\t*0.2},{\t}) -- (1.75,\t);
    }
    \mpsC{0}{1}{2}{colMPSBeta}
    \mpsB{0}{3}{colMPSBeta}
    \mpsA{0}{4}{colMPSBeta}
    \gridLine{1}{2}{1}{4}
    \gridLine{1}{-2-0.75}{1}{-4-0.75}
    \bdCircle{1}{3}{colU}
    \bCircle{1}{-3.75}{colUc}
    \node[anchor=west] at (1.75,4) {$j+1$};
    \node[anchor=west] at (1.75,3) {$j$};
    \node[anchor=west] at (1.75,2) {$j-1$};
    \node[anchor=west] at (1.75,1) {$j-2$};
    \foreach \t in {1,...,4}{\wedgeL{0}{\t}}
  \end{tikzpicture}=
  \begin{tikzpicture}[baseline={([yshift=-0.6ex]current bounding box.center)},scale=0.5]
    \mpsWire{0}{0.25}{0}{4.75}
    \foreach \t in {1,...,4}{
      \draw [thick,colLines,rounded corners=2] ({0.75},{-0.75-\t}) -- ({-0.5-\t*0.2},{-0.75-\t}) -- ({-0.5-\t*0.2},{\t}) -- (0.75,\t);
    }
    \mpsC{0}{3}{4}{colMPSBeta}
    \mpsB{0}{2}{colMPSBeta}
    \mpsA{0}{1}{colMPSBeta}
    \node[anchor=west] at (0.75,4) {$j+1$};
    \node[anchor=west] at (0.75,3) {$j$};
    \node[anchor=west] at (0.75,2) {$j-1$};
    \node[anchor=west] at (0.75,1) {$j-2$};
    \foreach \t in {1,...,4}{\wedgeL{0}{\t}}
  \end{tikzpicture},\qquad
  \begin{tikzpicture}[baseline={([yshift=-0.6ex]current bounding box.center)},scale=0.5]
    \mpsWire{0}{0}{0}{2.75}
    \foreach \t in {1,2}{
      \draw [thick,colLines,rounded corners=2] ({1.75},{-0.75-\t}) -- ({-0.5-\t*0.2},{-0.75-\t}) -- ({-0.5-\t*0.2},{\t}) -- (1.75,\t);
    }
    \mpsA{0}{2}{colMPSBeta}
    \mpsB{0}{1}{colMPSBeta}
    \mpsBvecV{0}{0}{colMPSBeta}
    \gridLine{1}{0}{1}{2}
    \gridLine{1}{-0.75}{1}{-2.75}
    \bdCircle{1}{1}{colU}
    \bCircle{1}{-1.75}{colUc}
    \vmpsW{1}{0}{colIst}{colLines}
    \vmpsW{1}{-0.75}{colIstC}{colLines}
    \node[anchor=west] at (1.75,2) {$j+1$};
    \node[anchor=west] at (1.75,1) {$j$};
    \foreach \t in {0,...,2}{\wedgeL{0}{\t}}
  \end{tikzpicture}=
  \begin{tikzpicture}[baseline={([yshift=-0.6ex]current bounding box.center)},scale=0.5]
    \mpsWire{0}{0}{0}{2.75}
    \foreach \t in {1,2}{
      \draw [thick,colLines,rounded corners=2] ({0.75},{-0.75-\t}) -- ({-0.5-\t*0.2},{-0.75-\t}) -- ({-0.5-\t*0.2},{\t}) -- (0.75,\t);
    }
    \mpsC{0}{1}{2}{colMPSBeta}
    \mpsBvecW{0}{0}{colMPSBeta}
    \node[anchor=west] at (0.75,2) {$j+1$};
    \node[anchor=west] at (0.75,1) {$j$};
    \foreach \t in {0,...,2}{\wedgeL{0}{\t}}
  \end{tikzpicture},\qquad
  \begin{tikzpicture}[baseline={([yshift=-0.6ex]current bounding box.center)},scale=0.5]
    \mpsWire{0}{0}{0}{3.75}
    \foreach \t in {1,...,3}{
      \draw [thick,colLines,rounded corners=2] ({1.75},{-0.75-\t}) -- ({-0.5-\t*0.2},{-0.75-\t}) -- ({-0.5-\t*0.2},{\t}) -- (1.75,\t);
    }
    \mpsA{0}{3}{colMPSBeta}
    \mpsB{0}{2}{colMPSBeta}
    \mpsA{0}{1}{colMPSBeta}
    \mpsBvecW{0}{0}{colMPSBeta}
    \gridLine{1}{0}{1}{3}
    \gridLine{1}{-0.75}{1}{-3.75}
    \bdCircle{1}{2}{colU}
    \bCircle{1}{-2.75}{colUc}
    \vmpsV{1}{0}{colIst}{colLines}
    \vmpsV{1}{-0.75}{colIstC}{colLines}
    \node[anchor=west] at (1.75,3) {$j+1$};
    \node[anchor=west] at (1.75,2) {$j$};
    \node[anchor=west] at (1.75,1) {$j-1$};
    \foreach \t in {0,...,3}{\wedgeL{0}{\t}}
  \end{tikzpicture}=
  \begin{tikzpicture}[baseline={([yshift=-0.6ex]current bounding box.center)},scale=0.5]
    \mpsWire{0}{0}{0}{3.75}
    \foreach \t in {1,...,3}{
      \draw [thick,colLines,rounded corners=2] ({0.75},{-0.75-\t}) -- ({-0.5-\t*0.2},{-0.75-\t}) -- ({-0.5-\t*0.2},{\t}) -- (0.75,\t);
    }
    \mpsC{0}{2}{3}{colMPSBeta}
    \mpsB{0}{1}{colMPSBeta}
    \mpsBvecV{0}{0}{colMPSBeta}
    \node[anchor=west] at (0.75,3) {$j+1$};
    \node[anchor=west] at (0.75,2) {$j$};
    \node[anchor=west] at (0.75,1) {$j-1$};
    \foreach \t in {0,...,3}{\wedgeL{0}{\t}}
  \end{tikzpicture},\\
  \begin{tikzpicture}[baseline={([yshift=-0.6ex]current bounding box.center)},scale=0.5]
    \mpsWire{0}{0}{0}{3.75}
    \foreach \t in {1,...,3}{
      \draw [thick,colLines,rounded corners=2] ({1.75},{-0.75-\t}) -- ({-0.5-\t*0.2},{-0.75-\t}) -- ({-0.5-\t*0.2},{\t}) -- (1.75,\t);
    }
    \draw [thick,colLines,rounded corners=2] (1,2) -- (1,4) -- (-1.5,4) -- (-1.5,-4.75) -- (1,-4.75) -- (1,-2.75);
    \mpsBvec{0}{3.75}
    \mpsB{0}{3}{colMPSBeta}
    \mpsC{0}{1}{2}{colMPSBeta}
    \bdCircle{1}{3}{colU}
    \bCircle{1}{-3.75}{colUc}
    \node[anchor=west] at (1.75,3) {$j+1$};
    \node[anchor=west] at (1.75,2) {$j$};
    \node[anchor=west] at (1.75,1) {$j-1$};
    \foreach \t in {1,...,3}{\wedgeL{0}{\t}}
  \end{tikzpicture}=
  \begin{tikzpicture}[baseline={([yshift=-0.6ex]current bounding box.center)},scale=0.5]
    \draw[ultra thick,white] (-0.5,-3.75-0.75) -- (0.5,-4.5);
    \mpsWire{0}{0}{0}{3.75}
    \foreach \t in {1,...,3}{
      \draw [thick,colLines,rounded corners=2] ({0.75},{-0.75-\t}) -- ({-0.5-\t*0.2},{-0.75-\t}) -- ({-0.5-\t*0.2},{\t}) -- (0.75,\t);
    }
    \mpsBvec{0}{3.75}
    \mpsA{0}{3}{colMPSBeta}
    \mpsB{0}{2}{colMPSBeta}
    \mpsA{0}{1}{colMPSBeta}
    \node[anchor=west] at (0.75,3) {$j+1$};
    \node[anchor=west] at (0.75,2) {$j$};
    \node[anchor=west] at (0.75,1) {$j-1$};
    \foreach \t in {1,...,3}{\wedgeL{0}{\t}}
  \end{tikzpicture},\qquad
  \begin{tikzpicture}[baseline={([yshift=-0.6ex]current bounding box.center)},scale=0.5]
    \mpsWire{0}{0}{0}{2.75}
    \foreach \t in {1,...,2}{
      \draw [thick,colLines,rounded corners=2] ({1.75},{-0.75-\t}) -- ({-0.5-\t*0.2},{-0.75-\t}) -- ({-0.5-\t*0.2},{\t}) -- (1.75,\t);
    }
    \draw [thick,colLines,rounded corners=2] (1,2) -- (1,3) -- (-1.5,3) -- (-1.5,-3.75) -- (1,-3.75) -- (1,-2.75);
    \mpsBvec{0}{2.75}
    \mpsC{0}{1}{2}{colMPSBeta}
    \node[anchor=west] at (1.75,2) {$j+1$};
    \node[anchor=west] at (1.75,1) {$j$};
    \foreach \t in {1,...,2}{\wedgeL{0}{\t}}
  \end{tikzpicture}=
  \begin{tikzpicture}[baseline={([yshift=-0.6ex]current bounding box.center)},scale=0.5]
    \draw[ultra thick,white] (-0.5,-3.5) -- (0.5,-3.5);
    \mpsWire{0}{0}{0}{2.75}
    \foreach \t in {1,...,2}{
      \draw [thick,colLines,rounded corners=2] ({0.75},{-0.75-\t}) -- ({-0.5-\t*0.2},{-0.75-\t}) -- ({-0.5-\t*0.2},{\t}) -- (0.75,\t);
    }
    \mpsBvec{0}{2.75}
    \mpsB{0}{2}{colMPSBeta}
    \mpsA{0}{1}{colMPSBeta}
    \node[anchor=west] at (0.75,2) {$j+1$};
    \node[anchor=west] at (0.75,1) {$j$};
    \foreach \t in {1,...,2}{\wedgeL{0}{\t}}
  \end{tikzpicture}.
  \end{gathered}
\end{equation}
Using these relations --- and assuming $|A|>2t$ --- we can rewrite Eq.~\eqref{eq:densityMatrixTN} as
\begin{equation}
  Z_{A}(\underline{\alpha})= z_{-}(\underline{\alpha}) z_{+}(\underline{\alpha}),\quad
  \text{with}\quad
  z_{-}(\underline{\alpha})=
\begin{tikzpicture}[baseline={([yshift=-0.6ex]current bounding box.center)},scale=0.5]
    \def\X{1}
    \def\Y{6}
    \foreach \t in {1,...,\Y}
    {
      \draw [thick,colLines,rounded corners=2] ({-0.5-\t*0.2},{\t}) rectangle ({2.5+0.2*\t},{-0.75-\t});
    }
    \draw [thick,colLines,rounded corners=2] (1,0) -- (1,\Y+1.25) -- (4.25,\Y+1.25) -- (4.25,-\Y-2) -- (1,-\Y-2) -- (1,-0.75);
    \foreach \x in {1}{
      \vmpsV{\x}{0}{colIst}{colLines}
      \vmpsV{\x}{-0.75}{colIstC}{colLines}
    }
    \mpsWire{0}{0}{0}{6.75}
    \mpsWire{2}{0}{2}{6.75}
    \mpsBvecW{0}{0}{colMPS}
    \mpsBvecW{2}{0}{colMPSBeta}
    \mpsBvec{0}{6.75}
    \mpsBvec{2}{6.75}
    \foreach \t in {1,3,...,\Y}
    {
      \bdLCircle{1}{1+\t}{colU}
      \bCircle{1}{-1.75-\t}{colUc}
      \mpsA{0}{\t}{colMPS}
      \mpsB{0}{\t+1}{colMPS}
      \mpsA{2}{\t}{colMPSBeta}
      \mpsB{2}{\t+1}{colMPSBeta}
    }
    \foreach \t in {0,...,\Y}{\wedgeR{2}{\t}}
    \node[anchor=west] at (4.25,1) {$1$};
    \node[anchor=west] at (4.25,2) {$2$};
    \node[anchor=west] at (4.25,3) {$3$};
    \node[anchor=west] at (4.25,4.5) {$\vdots$};
    \node[anchor=west] at (4.25,6) {$t$};
  \end{tikzpicture},\qquad
  z_{-}(\underline{\alpha},\underline{\beta})=
\begin{tikzpicture}[baseline={([yshift=-0.6ex]current bounding box.center)},scale=0.5]
    \def\X{1}
    \def\Y{6}
    \foreach \t in {1,...,\Y}
    {
      \draw [thick,colLines,rounded corners=2] ({-0.5-\t*0.2},{\t}) rectangle ({2.5+0.2*\t},{-0.75-\t});
    }
    \draw [thick,colLines,rounded corners=2] (1,0) -- (1,\Y+1.25) -- (-2.25,\Y+1.25) -- (-2.25,-\Y-2) -- (1,-\Y-2) -- (1,-0.75);
    \foreach \x in {1}{
      \vmpsW{\x}{0}{colIst}{colLines}
      \vmpsW{\x}{-0.75}{colIstC}{colLines}
    }
    \mpsWire{0}{0}{0}{6.75}
    \mpsWire{2}{0}{2}{6.75}
    \mpsBvecV{0}{0}{colMPSBeta}
    \mpsBvecV{2}{0}{colMPS}
    \mpsBvec{0}{6.75}
    \mpsBvec{2}{6.75}
    \foreach \t in {1,3,...,\Y}
    {
      \bdRCircle{1}{\t}{colU}
      \bCircle{1}{-0.75-\t}{colUc}
      \mpsB{2}{\t}{colMPS}
      \mpsA{2}{\t+1}{colMPS}
      \mpsB{0}{\t}{colMPSBeta}
      \mpsA{0}{\t+1}{colMPSBeta}
    }
    \foreach \t in {0,...,\Y}{\wedgeL{0}{\t}}
    \node[anchor=west] at (3.75,1) {$1$};
    \node[anchor=west] at (3.75,2) {$2$};
    \node[anchor=west] at (3.75,3) {$3$};
    \node[anchor=west] at (3.75,4.5) {$\vdots$};
    \node[anchor=west] at (3.75,6) {$t$};
  \end{tikzpicture}.
\end{equation}
To finally obtain the form in Eq.~\eqref{eq:R54ZwithFPs} we note that we are able to absorb the boundary layer of local gates into fixed points with the same bond dimension and modified parametrisation through the following set of local relations  
\begin{equation}
  \begin{gathered}
  \begin{tikzpicture}[baseline={([yshift=-0.6ex]current bounding box.center)},scale=0.5]
    \mpsWire{0}{0.25}{0}{4.75}
    \foreach \t in {1,...,4}{
      \draw [thick,colLines,rounded corners=2] ({1.75},{-0.75-\t}) -- ({-0.5-\t*0.2},{-0.75-\t}) -- ({-0.5-\t*0.2},{\t}) -- (1.75,\t);
    }
    \mpsC{0}{1}{2}{colMPSBetaB}
    \mpsB{0}{3}{colMPSBeta}
    \mpsA{0}{4}{colMPSBeta}
    \gridLine{1}{2}{1}{4}
    \gridLine{1}{-2-0.75}{1}{-4-0.75}
    \bdRCircle{1}{3}{colU}
    \bCircle{1}{-3.75}{colUc}
    \node[anchor=west] at (1.75,4) {$j+1$};
    \node[anchor=west] at (1.75,3) {$j$};
    \node[anchor=west] at (1.75,2) {$j-1$};
    \node[anchor=west] at (1.75,1) {$j-2$};
    \foreach \t in {1,...,4}{\wedgeL{0}{\t}}
  \end{tikzpicture}=
  \begin{tikzpicture}[baseline={([yshift=-0.6ex]current bounding box.center)},scale=0.5]
    \mpsWire{0}{0.25}{0}{4.75}
    \foreach \t in {1,...,4}{
      \draw [thick,colLines,rounded corners=2] ({0.75},{-0.75-\t}) -- ({-0.5-\t*0.2},{-0.75-\t}) -- ({-0.5-\t*0.2},{\t}) -- (0.75,\t);
    }
    \mpsC{0}{3}{4}{colMPSBetaB}
    \mpsB{0}{2}{colMPSBetaB}
    \mpsA{0}{1}{colMPSBetaB}
    \node[anchor=west] at (0.75,4) {$j+1$};
    \node[anchor=west] at (0.75,3) {$j$};
    \node[anchor=west] at (0.75,2) {$j-1$};
    \node[anchor=west] at (0.75,1) {$j-2$};
    \foreach \t in {1,...,4}{\wedgeL{0}{\t}}
  \end{tikzpicture},\qquad
  \begin{tikzpicture}[baseline={([yshift=-0.6ex]current bounding box.center)},scale=0.5]
    \mpsWire{0}{0}{0}{2.75}
    \foreach \t in {1,2}{
      \draw [thick,colLines,rounded corners=2] ({1.75},{-0.75-\t}) -- ({-0.5-\t*0.2},{-0.75-\t}) -- ({-0.5-\t*0.2},{\t}) -- (1.75,\t);
    }
    \mpsA{0}{2}{colMPSBeta}
    \mpsB{0}{1}{colMPSBeta}
    \mpsBvecV{0}{0}{colMPSBeta}
    \gridLine{1}{0}{1}{2}
    \gridLine{1}{-0.75}{1}{-2.75}
    \bdRCircle{1}{1}{colU}
    \bCircle{1}{-1.75}{colUc}
    \vmpsW{1}{0}{colIst}{colLines}
    \vmpsW{1}{-0.75}{colIstC}{colLines}
    \node[anchor=west] at (1.75,2) {$j+1$};
    \node[anchor=west] at (1.75,1) {$j$};
    \foreach \t in {0,...,2}{\wedgeL{0}{\t}}
  \end{tikzpicture}=
  \begin{tikzpicture}[baseline={([yshift=-0.6ex]current bounding box.center)},scale=0.5]
    \mpsWire{0}{0}{0}{2.75}
    \foreach \t in {1,2}{
      \draw [thick,colLines,rounded corners=2] ({0.75},{-0.75-\t}) -- ({-0.5-\t*0.2},{-0.75-\t}) -- ({-0.5-\t*0.2},{\t}) -- (0.75,\t);
    }
    \mpsC{0}{1}{2}{colMPSBetaB}
    \mpsBvecW{0}{0}{colMPSBetaB}
    \node[anchor=west] at (0.75,2) {$j+1$};
    \node[anchor=west] at (0.75,1) {$j$};
    \foreach \t in {0,...,2}{\wedgeL{0}{\t}}
  \end{tikzpicture},\qquad
  \begin{tikzpicture}[baseline={([yshift=-0.6ex]current bounding box.center)},scale=0.5]
    \mpsWire{0}{0}{0}{3.75}
    \foreach \t in {1,...,3}{
      \draw [thick,colLines,rounded corners=2] ({1.75},{-0.75-\t}) -- ({-0.5-\t*0.2},{-0.75-\t}) -- ({-0.5-\t*0.2},{\t}) -- (1.75,\t);
    }
    \mpsA{0}{3}{colMPSBeta}
    \mpsB{0}{2}{colMPSBeta}
    \mpsA{0}{1}{colMPSBeta}
    \mpsBvecW{0}{0}{colMPSBeta}
    \gridLine{1}{0}{1}{3}
    \gridLine{1}{-0.75}{1}{-3.75}
    \bdRCircle{1}{2}{colU}
    \bCircle{1}{-2.75}{colUc}
    \vmpsV{1}{0}{colIst}{colLines}
    \vmpsV{1}{-0.75}{colIstC}{colLines}
    \node[anchor=west] at (1.75,3) {$j+1$};
    \node[anchor=west] at (1.75,2) {$j$};
    \node[anchor=west] at (1.75,1) {$j-1$};
    \foreach \t in {0,...,3}{\wedgeL{0}{\t}}
  \end{tikzpicture}=
  \begin{tikzpicture}[baseline={([yshift=-0.6ex]current bounding box.center)},scale=0.5]
    \mpsWire{0}{0}{0}{3.75}
    \foreach \t in {1,...,3}{
      \draw [thick,colLines,rounded corners=2] ({0.75},{-0.75-\t}) -- ({-0.5-\t*0.2},{-0.75-\t}) -- ({-0.5-\t*0.2},{\t}) -- (0.75,\t);
    }
    \mpsC{0}{2}{3}{colMPSBetaB}
    \mpsB{0}{1}{colMPSBetaB}
    \mpsBvecV{0}{0}{colMPSBetaB}
    \node[anchor=west] at (0.75,3) {$j+1$};
    \node[anchor=west] at (0.75,2) {$j$};
    \node[anchor=west] at (0.75,1) {$j-1$};
    \foreach \t in {0,...,3}{\wedgeL{0}{\t}}
  \end{tikzpicture},\\
  \begin{tikzpicture}[baseline={([yshift=-0.6ex]current bounding box.center)},scale=0.5]
    \mpsWire{0}{0}{0}{3.75}
    \foreach \t in {1,...,3}{
      \draw [thick,colLines,rounded corners=2] ({1.75},{-0.75-\t}) -- ({-0.5-\t*0.2},{-0.75-\t}) -- ({-0.5-\t*0.2},{\t}) -- (1.75,\t);
    }
    \draw [thick,colLines,rounded corners=2] (1,2) -- (1,4) -- (-1.5,4) -- (-1.5,-4.75) -- (1,-4.75) -- (1,-2.75);
    \mpsBvec{0}{3.75}
    \mpsB{0}{3}{colMPSBeta}
    \mpsC{0}{1}{2}{colMPSBetaB}
    \bdRCircle{1}{3}{colU}
    \bCircle{1}{-3.75}{colUc}
    \node[anchor=west] at (1.75,3) {$j+1$};
    \node[anchor=west] at (1.75,2) {$j$};
    \node[anchor=west] at (1.75,1) {$j-1$};
    \foreach \t in {1,...,3}{\wedgeL{0}{\t}}
  \end{tikzpicture}=
  \begin{tikzpicture}[baseline={([yshift=-0.6ex]current bounding box.center)},scale=0.5]
    \draw[ultra thick,white] (-0.5,-3.75-0.75) -- (0.5,-4.5);
    \mpsWire{0}{0}{0}{3.75}
    \foreach \t in {1,...,3}{
      \draw [thick,colLines,rounded corners=2] ({0.75},{-0.75-\t}) -- ({-0.5-\t*0.2},{-0.75-\t}) -- ({-0.5-\t*0.2},{\t}) -- (0.75,\t);
    }
    \mpsBvec{0}{3.75}
    \mpsA{0}{3}{colMPSBetaB}
    \mpsB{0}{2}{colMPSBetaB}
    \mpsA{0}{1}{colMPSBetaB}
    \node[anchor=west] at (0.75,3) {$j+1$};
    \node[anchor=west] at (0.75,2) {$j$};
    \node[anchor=west] at (0.75,1) {$j-1$};
    \foreach \t in {1,...,3}{\wedgeL{0}{\t}}
  \end{tikzpicture},\qquad
  \begin{tikzpicture}[baseline={([yshift=-0.6ex]current bounding box.center)},scale=0.5]
    \mpsWire{0}{0}{0}{2.75}
    \foreach \t in {1,...,2}{
      \draw [thick,colLines,rounded corners=2] ({1.75},{-0.75-\t}) -- ({-0.5-\t*0.2},{-0.75-\t}) -- ({-0.5-\t*0.2},{\t}) -- (1.75,\t);
    }
    \draw [thick,colLines,rounded corners=2] (1,2) -- (1,3) -- (-1.5,3) -- (-1.5,-3.75) -- (1,-3.75) -- (1,-2.75);
    \mpsBvec{0}{2.75}
    \mpsC{0}{1}{2}{colMPSBetaB}
    \node[anchor=west] at (1.75,2) {$j+1$};
    \node[anchor=west] at (1.75,1) {$j$};
    \foreach \t in {1,...,2}{\wedgeL{0}{\t}}
  \end{tikzpicture}=
  \begin{tikzpicture}[baseline={([yshift=-0.6ex]current bounding box.center)},scale=0.5]
    \draw[ultra thick,white] (-0.5,-3.5) -- (0.5,-3.5);
    \mpsWire{0}{0}{0}{2.75}
    \foreach \t in {1,...,2}{
      \draw [thick,colLines,rounded corners=2] ({0.75},{-0.75-\t}) -- ({-0.5-\t*0.2},{-0.75-\t}) -- ({-0.5-\t*0.2},{\t}) -- (0.75,\t);
    }
    \mpsBvec{0}{2.75}
    \mpsB{0}{2}{colMPSBetaB}
    \mpsA{0}{1}{colMPSBetaB}
    \node[anchor=west] at (0.75,2) {$j+1$};
    \node[anchor=west] at (0.75,1) {$j$};
    \foreach \t in {1,...,2}{\wedgeL{0}{\t}}
  \end{tikzpicture}.
  \end{gathered}
\end{equation}

\subsubsection{Fixed-point tensors}
For completeness we now explicitly report the ``bulk'' fixed-point tensors introduced above 
\begin{equation}
\begin{aligned}
  \begin{tikzpicture}[baseline={([yshift=-0.6ex]current bounding box.center)},scale=0.5]
    \mpsWire{0}{-0.75}{0}{0.75}
    \gridLine{-0.75}{0}{0.75}{0}
    \mpsA{0}{0}{colMPSBeta}
    \wedgeL{0}{0}
    \node at (-1,0) {$0$};
    \node at (1,0) {$0$};
  \end{tikzpicture}&=
  \begin{bmatrix}
    1-\vartheta & 1-\vartheta & -(1-\vartheta) \\
    \Pi_j \vartheta & \Pi_j \vartheta & \Pi_j (1-\vartheta) \\
    \vartheta & - \frac{\vartheta^2}{1-\vartheta} & - \vartheta
  \end{bmatrix},&\qquad
  \begin{tikzpicture}[baseline={([yshift=-0.6ex]current bounding box.center)},scale=0.5]
    \mpsWire{0}{-0.75}{0}{0.75}
    \gridLine{-0.75}{0}{0.75}{0}
    \mpsA{0}{0}{colMPSBeta}
    \wedgeL{0}{0}
    \node at (-1,0) {$1$};
    \node at (1,0) {$1$};
  \end{tikzpicture}&=
  \begin{bmatrix}
    0 & 1 & 0 \\
    \Pi_{j-1} & 0 & 0 \\
    0 & 0 & 0
  \end{bmatrix},\\
  \begin{tikzpicture}[baseline={([yshift=-0.6ex]current bounding box.center)},scale=0.5]
    \mpsWire{0}{-0.75}{0}{0.75}
    \gridLine{-0.75}{0}{0.75}{0}
    \mpsA{0}{0}{colMPSBeta}
    \wedgeL{0}{0}
    \node at (-1,0) {$0$};
    \node at (1,0) {$1$};
  \end{tikzpicture}&=
  \begin{bmatrix}
    0 & 1-\vartheta & - (1-\vartheta) \\
    \mathrm{e}^{\alpha_j}\Pi_j \vartheta& 0 & 0 \\
    \mathrm{e}^{\alpha_j}\vartheta & 0 & 0 
  \end{bmatrix},&
  \begin{tikzpicture}[baseline={([yshift=-0.6ex]current bounding box.center)},scale=0.5]
    \mpsWire{0}{-0.75}{0}{0.75}
    \gridLine{-0.75}{0}{0.75}{0}
    \mpsA{0}{0}{colMPSBeta}
    \wedgeL{0}{0}
    \node at (-1,0) {$1$};
    \node at (1,0) {$0$};
  \end{tikzpicture}&
  \xleftrightarrow{\alpha_j\leftrightarrow \beta_j}
  \begin{tikzpicture}[baseline={([yshift=-0.6ex]current bounding box.center)},scale=0.5]
    \mpsWire{0}{-0.75}{0}{0.75}
    \gridLine{-0.75}{0}{0.75}{0}
    \mpsA{0}{0}{colMPSBeta}
    \wedgeL{0}{0}
    \node at (-1,0) {$0$};
    \node at (1,0) {$1$};
  \end{tikzpicture}
  \\
  \begin{tikzpicture}[baseline={([yshift=-0.6ex]current bounding box.center)},scale=0.5]
    \mpsWire{0}{-0.75}{0}{0.75}
    \gridLine{-0.75}{0}{0.75}{0}
    \mpsB{0}{0}{colMPSBeta}
    \wedgeL{0}{0}
    \node at (-1,0) {$s$};
    \node at (1,0) {$b$};
  \end{tikzpicture}&=
  \begin{bmatrix}
    \delta_{s,0} & 0 & 0 \\ 0 & \delta_{s,1} & 0 \\ 0 & 0 & \delta_{s,1} \Pi_{j-1}
  \end{bmatrix},&
  \begin{tikzpicture}[baseline={([yshift=-0.6ex]current bounding box.center)},scale=0.5]
    \mpsWire{0}{0}{0}{0.75}
    \mpsBvecW{0}{0}{colMPSBeta}
    \wedgeL{0}{0}
  \end{tikzpicture}=
  \begin{bmatrix} 1 \\ 0 \\ 0 \end{bmatrix},\qquad
  \begin{tikzpicture}[baseline={([yshift=-0.6ex]current bounding box.center)},scale=0.5]
    \mpsWire{0}{0}{0}{0.75}
    \mpsBvecV{0}{0}{colMPSBeta}
    \wedgeL{0}{0}
  \end{tikzpicture}&=
  \begin{bmatrix} 1-\vartheta \\ \Pi_0 \vartheta \\ -\frac{\vartheta^2}{1-\vartheta}\end{bmatrix},\qquad
  \begin{tikzpicture}[baseline={([yshift=-0.6ex]current bounding box.center)},scale=0.5]
    \mpsWire{0}{-0.75}{0}{0}
    \mpsBvec{0}{0}
  \end{tikzpicture}=\begin{bmatrix} 1 & 1 & 0 \end{bmatrix},
\end{aligned}
\end{equation}
where all the matrices are assumed to be applied at time-step $j$. The auxiliary two-time-step tensors are given in Sec.~\ref{sec:Cmatrices}. The tensors corresponding to the right fixed-point have an analogous form, and for the trivial choice of $\underline{\alpha}$ we recover the usual fixed-point tensors
\begin{equation}
  \left\{
  \begin{tikzpicture}[baseline={([yshift=-0.6ex]current bounding box.center)},scale=0.5]
    \mpsWire{0}{-0.75}{0}{0.75}
    \gridLine{-0.75}{0}{0.75}{0}
    \mpsA{0}{0}{colMPSBeta}
    \wedgeL{0}{0}
  \end{tikzpicture},
  \begin{tikzpicture}[baseline={([yshift=-0.6ex]current bounding box.center)},scale=0.5]
    \mpsWire{0}{-0.75}{0}{0.75}
    \gridLine{-0.75}{0}{0.75}{0}
    \mpsB{0}{0}{colMPSBeta}
    \wedgeL{0}{0}
  \end{tikzpicture},
  \begin{tikzpicture}[baseline={([yshift=-0.6ex]current bounding box.center)},scale=0.5]
    \mpsWire{0}{-1.75}{0}{0.75}
    \gridLine{-0.75}{0}{0.75}{0}
    \gridLine{-0.75}{-1}{0.75}{-1}
    \mpsC{0}{-1}{0}{colMPSBeta}
    \wedgeL{0}{0}
    \wedgeL{0}{-1}
  \end{tikzpicture},
  \begin{tikzpicture}[baseline={([yshift=-0.6ex]current bounding box.center)},scale=0.5]
    \mpsWire{0}{0}{0}{0.75}
    \mpsBvecW{0}{0}{colMPSBeta}
    \wedgeL{0}{0}
  \end{tikzpicture},
  \begin{tikzpicture}[baseline={([yshift=-0.6ex]current bounding box.center)},scale=0.5]
    \mpsWire{0}{0}{0}{0.75}
    \mpsBvecV{0}{0}{colMPSBeta}
    \wedgeL{0}{0}
  \end{tikzpicture}\right\}
  \xleftrightarrow[\Pi_j\leftrightarrow \frac{1}{\Pi_j}]{\alpha_j \leftrightarrow -\alpha_j}
  \left\{
  \begin{tikzpicture}[baseline={([yshift=-0.6ex]current bounding box.center)},scale=0.5]
    \mpsWire{0}{-0.75}{0}{0.75}
    \gridLine{-0.75}{0}{0.75}{0}
    \mpsA{0}{0}{colMPSBeta}
    \wedgeR{0}{0}
  \end{tikzpicture},
  \begin{tikzpicture}[baseline={([yshift=-0.6ex]current bounding box.center)},scale=0.5]
    \mpsWire{0}{-0.75}{0}{0.75}
    \gridLine{-0.75}{0}{0.75}{0}
    \mpsB{0}{0}{colMPSBeta}
    \wedgeR{0}{0}
  \end{tikzpicture},
  \begin{tikzpicture}[baseline={([yshift=-0.6ex]current bounding box.center)},scale=0.5]
    \mpsWire{0}{-1.75}{0}{0.75}
    \gridLine{-0.75}{0}{0.75}{0}
    \gridLine{-0.75}{-1}{0.75}{-1}
    \mpsC{0}{-1}{0}{colMPSBeta}
    \wedgeR{0}{0}
    \wedgeR{0}{-1}
  \end{tikzpicture},
  \begin{tikzpicture}[baseline={([yshift=-0.6ex]current bounding box.center)},scale=0.5]
    \mpsWire{0}{0}{0}{0.75}
    \mpsBvecW{0}{0}{colMPSBeta}
    \wedgeR{0}{0}
  \end{tikzpicture},
  \begin{tikzpicture}[baseline={([yshift=-0.6ex]current bounding box.center)},scale=0.5]
    \mpsWire{0}{0}{0}{0.75}
    \mpsBvecV{0}{0}{colMPSBeta}
    \wedgeR{0}{0}
  \end{tikzpicture}\right\}
\xrightarrow[\Pi_j\to 1]{\alpha_j,\beta_j \to 0}=
  \left\{
  \begin{tikzpicture}[baseline={([yshift=-0.6ex]current bounding box.center)},scale=0.5]
    \mpsWire{0}{-0.75}{0}{0.75}
    \gridLine{-0.75}{0}{0.75}{0}
    \mpsA{0}{0}{colMPS}
  \end{tikzpicture},
  \begin{tikzpicture}[baseline={([yshift=-0.6ex]current bounding box.center)},scale=0.5]
    \mpsWire{0}{-0.75}{0}{0.75}
    \gridLine{-0.75}{0}{0.75}{0}
    \mpsB{0}{0}{colMPS}
  \end{tikzpicture},
  \begin{tikzpicture}[baseline={([yshift=-0.6ex]current bounding box.center)},scale=0.5]
    \mpsWire{0}{-1.75}{0}{0.75}
    \gridLine{-0.75}{0}{0.75}{0}
    \gridLine{-0.75}{-1}{0.75}{-1}
    \mpsC{0}{-1}{0}{colMPS}
  \end{tikzpicture},
  \begin{tikzpicture}[baseline={([yshift=-0.6ex]current bounding box.center)},scale=0.5]
    \mpsWire{0}{0}{0}{0.75}
    \mpsBvecW{0}{0}{colMPS}
  \end{tikzpicture},
  \begin{tikzpicture}[baseline={([yshift=-0.6ex]current bounding box.center)},scale=0.5]
    \mpsWire{0}{0}{0}{0.75}
    \mpsBvecV{0}{0}{colMPS}
  \end{tikzpicture}\right\}.
\end{equation}

\subsubsection{Auxiliary two-timestep fixed-point tensors}\label{sec:Cmatrices}
Here we report the auxiliary two-time-step tensor, assuming that it is applied at time-steps labelled by $j$ and $j+1$ (the bottom one is $j$, the top one $j+1$)
\begin{equation}
  \begin{aligned}
  \begin{tikzpicture}[baseline={([yshift=-0.6ex]current bounding box.center)},scale=0.5]
    \mpsWire{0}{-1.75}{0}{0.75}
    \gridLine{-0.75}{0}{0.75}{0}
    \gridLine{-0.75}{-1}{0.75}{-1}
    \mpsC{0}{-1}{0}{colMPSBeta}
    \wedgeL{0}{0}
    \wedgeL{0}{-1}
    \node at (-1.25,0) {$0$};
    \node at (1.25,0) {$0$};
    \node at (-1.25,-1) {$0$};
    \node at (1.25,-1) {$0$};
  \end{tikzpicture}&=
\begin{tikzpicture}[baseline={([yshift=-0.6ex]current bounding box.center)},scale=0.5]
    \mpsWire{0}{-1.75}{0}{0.75}
    \gridLine{-0.75}{0}{0.75}{0}
    \gridLine{-0.75}{-1}{0.75}{-1}
    \mpsC{0}{-1}{0}{colMPSBetaB}
    \wedgeL{0}{0}
    \wedgeL{0}{-1}
    \node at (-1.25,0) {$0$};
    \node at (1.25,0) {$0$};
    \node at (-1.25,-1) {$0$};
    \node at (1.25,-1) {$0$};
  \end{tikzpicture}=
  \begin{bmatrix}
    (1-\vartheta)^2 & (1-\vartheta)^2 & -(1-\vartheta)^2\\
    \Pi_{j+1} \vartheta(1-\vartheta) & \Pi_{j+1} \vartheta(1-\vartheta)
    & -\Pi_{j+1}\vartheta(1-\vartheta)\\
    \vartheta(1-\vartheta) & -\vartheta^2 & \vartheta^2
  \end{bmatrix},\\
  \begin{tikzpicture}[baseline={([yshift=-0.6ex]current bounding box.center)},scale=0.5]
    \mpsWire{0}{-1.75}{0}{0.75}
    \gridLine{-0.75}{0}{0.75}{0}
    \gridLine{-0.75}{-1}{0.75}{-1}
    \mpsC{0}{-1}{0}{colMPSBeta}
    \wedgeL{0}{0}
    \wedgeL{0}{-1}
    \node at (-1.25,0) {$0$};
    \node at (1.25,0) {$0$};
    \node at (-1.25,-1) {$1$};
    \node at (1.25,-1) {$1$};
  \end{tikzpicture}&=
  \begin{tikzpicture}[baseline={([yshift=-0.6ex]current bounding box.center)},scale=0.5]
    \mpsWire{0}{-1.75}{0}{0.75}
    \gridLine{-0.75}{0}{0.75}{0}
    \gridLine{-0.75}{-1}{0.75}{-1}
    \mpsC{0}{-1}{0}{colMPSBetaB}
    \wedgeL{0}{0}
    \wedgeL{0}{-1}
    \node at (-1.25,0) {$0$};
    \node at (1.25,0) {$0$};
    \node at (-1.25,-1) {$1$};
    \node at (1.25,-1) {$1$};
  \end{tikzpicture}=
  \begin{bmatrix}
    0 & 1-\vartheta & 0 \\ 0 & \Pi_{j+1} \vartheta & 0 \\ 0 & 0 & \vartheta
  \end{bmatrix},\qquad
  \begin{tikzpicture}[baseline={([yshift=-0.6ex]current bounding box.center)},scale=0.5]
      \mpsWire{0}{-1.75}{0}{0.75}
      \gridLine{-0.75}{0}{0.75}{0}
      \gridLine{-0.75}{-1}{0.75}{-1}
      \mpsC{0}{-1}{0}{colMPSBeta}
      \wedgeL{0}{0}
      \wedgeL{0}{-1}
      \node at (-1.25,0) {$1$};
      \node at (1.25,0) {$1$};
      \node at (-1.25,-1) {$0$};
      \node at (1.25,-1) {$0$};
  \end{tikzpicture}=
  \begin{tikzpicture}[baseline={([yshift=-0.6ex]current bounding box.center)},scale=0.5]
    \mpsWire{0}{-1.75}{0}{0.75}
    \gridLine{-0.75}{0}{0.75}{0}
    \gridLine{-0.75}{-1}{0.75}{-1}
    \mpsC{0}{-1}{0}{colMPSBetaB}
    \wedgeL{0}{0}
    \wedgeL{0}{-1}
    \node at (-1.25,0) {$1$};
    \node at (1.25,0) {$1$};
    \node at (-1.25,-1) {$0$};
    \node at (1.25,-1) {$0$};
  \end{tikzpicture}=
  \begin{bmatrix}
    \Pi_j\vartheta & 0 & 0 \\
    0 & \Pi_{j}\vartheta & \Pi_j(1-\vartheta)\\
    0 & 0 & 0
  \end{bmatrix},\\
  \begin{tikzpicture}[baseline={([yshift=-0.6ex]current bounding box.center)},scale=0.5]
    \mpsWire{0}{-1.75}{0}{0.75}
    \gridLine{-0.75}{0}{0.75}{0}
    \gridLine{-0.75}{-1}{0.75}{-1}
    \mpsC{0}{-1}{0}{colMPSBeta}
    \wedgeL{0}{0}
    \wedgeL{0}{-1}
    \node at (-1.25,0) {$0$};
    \node at (1.25,0) {$0$};
    \node at (-1.25,-1) {$0$};
    \node at (1.25,-1) {$1$};
  \end{tikzpicture}&=
  \begin{tikzpicture}[baseline={([yshift=-0.6ex]current bounding box.center)},scale=0.5]
    \mpsWire{0}{-1.75}{0}{0.75}
    \gridLine{-0.75}{0}{0.75}{0}
    \gridLine{-0.75}{-1}{0.75}{-1}
    \mpsC{0}{-1}{0}{colMPSBetaB}
    \wedgeL{0}{0}
    \wedgeL{0}{-1}
    \node at (-1.25,0) {$0$};
    \node at (1.25,0) {$0$};
    \node at (-1.25,-1) {$0$};
    \node at (1.25,-1) {$1$};
  \end{tikzpicture}=
  \begin{tikzpicture}[baseline={([yshift=-0.6ex]current bounding box.center)},scale=0.5]
    \mpsWire{0}{-1.75}{0}{0.75}
    \gridLine{-0.75}{0}{0.75}{0}
    \gridLine{-0.75}{-1}{0.75}{-1}
    \mpsC{0}{-1}{0}{colMPSBeta}
    \wedgeL{0}{0}
    \wedgeL{0}{-1}
    \node at (-1.25,0) {$0$};
    \node at (1.25,0) {$0$};
    \node at (-1.25,-1) {$1$};
    \node at (1.25,-1) {$0$};
  \end{tikzpicture}=
  \begin{tikzpicture}[baseline={([yshift=-0.6ex]current bounding box.center)},scale=0.5]
    \mpsWire{0}{-1.75}{0}{0.75}
    \gridLine{-0.75}{0}{0.75}{0}
    \gridLine{-0.75}{-1}{0.75}{-1}
    \mpsC{0}{-1}{0}{colMPSBetaB}
    \wedgeL{0}{0}
    \wedgeL{0}{-1}
    \node at (-1.25,0) {$0$};
    \node at (1.25,0) {$0$};
    \node at (-1.25,-1) {$1$};
    \node at (1.25,-1) {$0$};
  \end{tikzpicture}=
  \begin{bmatrix}
    0 & (1-\vartheta)^2 & - (1-\vartheta)^2 \\
    0 & \Pi_{j+1} \vartheta (1-\vartheta)& -\Pi_{j+1}\vartheta(1-\vartheta) \\
    0 & -\vartheta^2 & \vartheta^2
  \end{bmatrix},\\
  \begin{tikzpicture}[baseline={([yshift=-0.6ex]current bounding box.center)},scale=0.5]
    \mpsWire{0}{-1.75}{0}{0.75}
    \gridLine{-0.75}{0}{0.75}{0}
    \gridLine{-0.75}{-1}{0.75}{-1}
    \mpsC{0}{-1}{0}{colMPSBeta}
    \wedgeL{0}{0}
    \wedgeL{0}{-1}
    \node at (-1.25,0) {$0$};
    \node at (1.25,0) {$1$};
    \node at (-1.25,-1) {$s$};
    \node at (1.25,-1) {$s$};
  \end{tikzpicture}&=
  \begin{tikzpicture}[baseline={([yshift=-0.6ex]current bounding box.center)},scale=0.5]
    \mpsWire{0}{-1.75}{0}{0.75}
    \gridLine{-0.75}{0}{0.75}{0}
    \gridLine{-0.75}{-1}{0.75}{-1}
    \mpsC{0}{-1}{0}{colMPSBetaB}
    \wedgeL{0}{0}
    \wedgeL{0}{-1}
    \node at (-1.25,0) {$0$};
    \node at (1.25,0) {$1$};
    \node at (-1.25,-1) {$s$};
    \node at (1.25,-1) {$s$};
  \end{tikzpicture}=
  \begin{tikzpicture}[baseline={([yshift=-0.6ex]current bounding box.center)},scale=0.5]
    \mpsWire{0}{-1.75}{0}{0.75}
    \gridLine{-0.75}{0}{0.75}{0}
    \gridLine{-0.75}{-1}{0.75}{-1}
    \mpsC{0}{-1}{0}{colMPSBeta}
    \wedgeL{0}{0}
    \wedgeL{0}{-1}
    \node at (-1.25,0) {$1$};
    \node at (1.25,0) {$0$};
    \node at (-1.25,-1) {$s$};
    \node at (1.25,-1) {$s$};
  \end{tikzpicture}=
  \begin{tikzpicture}[baseline={([yshift=-0.6ex]current bounding box.center)},scale=0.5]
    \mpsWire{0}{-1.75}{0}{0.75}
    \gridLine{-0.75}{0}{0.75}{0}
    \gridLine{-0.75}{-1}{0.75}{-1}
    \mpsC{0}{-1}{0}{colMPSBetaB}
    \wedgeL{0}{0}
    \wedgeL{0}{-1}
    \node at (-1.25,0) {$1$};
    \node at (1.25,0) {$0$};
    \node at (-1.25,-1) {$s$};
    \node at (1.25,-1) {$s$};
  \end{tikzpicture}=
  0,\\
  \begin{tikzpicture}[baseline={([yshift=-0.6ex]current bounding box.center)},scale=0.5]
      \mpsWire{0}{-1.75}{0}{0.75}
      \gridLine{-0.75}{0}{0.75}{0}
      \gridLine{-0.75}{-1}{0.75}{-1}
      \mpsC{0}{-1}{0}{colMPSBeta}
      \wedgeL{0}{0}
      \wedgeL{0}{-1}
      \node at (-1.25,0) {$0$};
      \node at (1.25,0) {$1$};
      \node at (-1.25,-1) {$1$};
      \node at (1.25,-1) {$0$};
  \end{tikzpicture}&=
  \begin{tikzpicture}[baseline={([yshift=-0.6ex]current bounding box.center)},scale=0.5]
    \mpsWire{0}{-1.75}{0}{0.75}
    \gridLine{-0.75}{0}{0.75}{0}
    \gridLine{-0.75}{-1}{0.75}{-1}
    \mpsC{0}{-1}{0}{colMPSBetaB}
    \wedgeL{0}{0}
    \wedgeL{0}{-1}
    \node at (-1.25,0) {$0$};
    \node at (1.25,0) {$1$};
    \node at (-1.25,-1) {$1$};
    \node at (1.25,-1) {$0$};
  \end{tikzpicture}=
  \mathrm{e}^{-\alpha_{j+1}}
\begin{tikzpicture}[baseline={([yshift=-0.6ex]current bounding box.center)},scale=0.5]
      \mpsWire{0}{-1.75}{0}{0.75}
      \gridLine{-0.75}{0}{0.75}{0}
      \gridLine{-0.75}{-1}{0.75}{-1}
      \mpsC{0}{-1}{0}{colMPSBeta}
      \wedgeL{0}{0}
      \wedgeL{0}{-1}
      \node at (-1.25,0) {$1$};
      \node at (1.25,0) {$0$};
      \node at (-1.25,-1) {$0$};
      \node at (1.25,-1) {$1$};
  \end{tikzpicture}=
  \mathrm{e}^{-\alpha_{j+1}}
  \begin{tikzpicture}[baseline={([yshift=-0.6ex]current bounding box.center)},scale=0.5]
    \mpsWire{0}{-1.75}{0}{0.75}
    \gridLine{-0.75}{0}{0.75}{0}
    \gridLine{-0.75}{-1}{0.75}{-1}
    \mpsC{0}{-1}{0}{colMPSBetaB}
    \wedgeL{0}{0}
    \wedgeL{0}{-1}
    \node at (-1.25,0) {$1$};
    \node at (1.25,0) {$0$};
    \node at (-1.25,-1) {$0$};
    \node at (1.25,-1) {$1$};
  \end{tikzpicture}=
  \begin{bmatrix}
    0 & 0 & 0 \\
    0 & \Pi_{j+1}\vartheta^2 & \Pi_{j+1} \vartheta (1-\vartheta)\\
    0 & \vartheta^2 & \vartheta(1-\vartheta)
  \end{bmatrix},\\
  \begin{tikzpicture}[baseline={([yshift=-0.6ex]current bounding box.center)},scale=0.5]
      \mpsWire{0}{-1.75}{0}{0.75}
      \gridLine{-0.75}{0}{0.75}{0}
      \gridLine{-0.75}{-1}{0.75}{-1}
      \mpsC{0}{-1}{0}{colMPSBeta}
      \wedgeL{0}{0}
      \wedgeL{0}{-1}
      \node at (-1.25,0) {$0$};
      \node at (1.25,0) {$1$};
      \node at (-1.25,-1) {$0$};
      \node at (1.25,-1) {$1$};
  \end{tikzpicture}&=
  \begin{tikzpicture}[baseline={([yshift=-0.6ex]current bounding box.center)},scale=0.5]
    \mpsWire{0}{-1.75}{0}{0.75}
    \gridLine{-0.75}{0}{0.75}{0}
    \gridLine{-0.75}{-1}{0.75}{-1}
    \mpsC{0}{-1}{0}{colMPSBetaB}
    \wedgeL{0}{0}
    \wedgeL{0}{-1}
    \node at (-1.25,0) {$0$};
    \node at (1.25,0) {$1$};
    \node at (-1.25,-1) {$0$};
    \node at (1.25,-1) {$1$};
  \end{tikzpicture}=
  \mathrm{e}^{-\alpha_j-\alpha_{j+1}}
\begin{tikzpicture}[baseline={([yshift=-0.6ex]current bounding box.center)},scale=0.5]
      \mpsWire{0}{-1.75}{0}{0.75}
      \gridLine{-0.75}{0}{0.75}{0}
      \gridLine{-0.75}{-1}{0.75}{-1}
      \mpsC{0}{-1}{0}{colMPSBeta}
      \wedgeL{0}{0}
      \wedgeL{0}{-1}
      \node at (-1.25,0) {$1$};
      \node at (1.25,0) {$0$};
      \node at (-1.25,-1) {$1$};
      \node at (1.25,-1) {$0$};
  \end{tikzpicture}=
  \mathrm{e}^{-\alpha_{j+1}}
  \begin{tikzpicture}[baseline={([yshift=-0.6ex]current bounding box.center)},scale=0.5]
    \mpsWire{0}{-1.75}{0}{0.75}
    \gridLine{-0.75}{0}{0.75}{0}
    \gridLine{-0.75}{-1}{0.75}{-1}
    \mpsC{0}{-1}{0}{colMPSBetaB}
    \wedgeL{0}{0}
    \wedgeL{0}{-1}
    \node at (-1.25,0) {$1$};
    \node at (1.25,0) {$0$};
    \node at (-1.25,-1) {$1$};
    \node at (1.25,-1) {$0$};
  \end{tikzpicture}=
  \begin{bmatrix}
    0 & 0 & 0 \\
    \Pi_{j+1} \vartheta(1-\vartheta) & 0 & 0 \\
    \vartheta(1-\vartheta) & 0 & 0
  \end{bmatrix},\\
  \begin{tikzpicture}[baseline={([yshift=-0.6ex]current bounding box.center)},scale=0.5]
      \mpsWire{0}{-1.75}{0}{0.75}
      \gridLine{-0.75}{0}{0.75}{0}
      \gridLine{-0.75}{-1}{0.75}{-1}
      \mpsC{0}{-1}{0}{colMPSBeta}
      \wedgeL{0}{0}
      \wedgeL{0}{-1}
      \node at (-1.25,0) {$1$};
      \node at (1.25,0) {$1$};
      \node at (-1.25,-1) {$0$};
      \node at (1.25,-1) {$1$};
  \end{tikzpicture}&=
  \begin{tikzpicture}[baseline={([yshift=-0.6ex]current bounding box.center)},scale=0.5]
    \mpsWire{0}{-1.75}{0}{0.75}
    \gridLine{-0.75}{0}{0.75}{0}
    \gridLine{-0.75}{-1}{0.75}{-1}
    \mpsC{0}{-1}{0}{colMPSBetaB}
    \wedgeL{0}{0}
    \wedgeL{0}{-1}
    \node at (-1.25,0) {$1$};
    \node at (1.25,0) {$1$};
    \node at (-1.25,-1) {$0$};
    \node at (1.25,-1) {$1$};
  \end{tikzpicture}=
  \mathrm{e}^{-\alpha_j}
\begin{tikzpicture}[baseline={([yshift=-0.6ex]current bounding box.center)},scale=0.5]
      \mpsWire{0}{-1.75}{0}{0.75}
      \gridLine{-0.75}{0}{0.75}{0}
      \gridLine{-0.75}{-1}{0.75}{-1}
      \mpsC{0}{-1}{0}{colMPSBeta}
      \wedgeL{0}{0}
      \wedgeL{0}{-1}
      \node at (-1.25,0) {$1$};
      \node at (1.25,0) {$1$};
      \node at (-1.25,-1) {$1$};
      \node at (1.25,-1) {$0$};
  \end{tikzpicture}=
  \begin{tikzpicture}[baseline={([yshift=-0.6ex]current bounding box.center)},scale=0.5]
    \mpsWire{0}{-1.75}{0}{0.75}
    \gridLine{-0.75}{0}{0.75}{0}
    \gridLine{-0.75}{-1}{0.75}{-1}
    \mpsC{0}{-1}{0}{colMPSBetaB}
    \wedgeL{0}{0}
    \wedgeL{0}{-1}
    \node at (-1.25,0) {$1$};
    \node at (1.25,0) {$1$};
    \node at (-1.25,-1) {$1$};
    \node at (1.25,-1) {$0$};
  \end{tikzpicture}=
  \begin{bmatrix}
    \Pi_j\vartheta & 0 & 0 \\
    0 & 0 & 0 \\
    0 & 0 & 0
  \end{bmatrix},\\
\begin{tikzpicture}[baseline={([yshift=-0.6ex]current bounding box.center)},scale=0.5]
      \mpsWire{0}{-1.75}{0}{0.75}
      \gridLine{-0.75}{0}{0.75}{0}
      \gridLine{-0.75}{-1}{0.75}{-1}
      \mpsC{0}{-1}{0}{colMPSBeta}
      \wedgeL{0}{0}
      \wedgeL{0}{-1}
      \node at (-1.25,0) {$1$};
      \node at (1.25,0) {$1$};
      \node at (-1.25,-1) {$1$};
      \node at (1.25,-1) {$1$};
\end{tikzpicture}&=
\mathrm{e}^{\alpha_j}
  \begin{tikzpicture}[baseline={([yshift=-0.6ex]current bounding box.center)},scale=0.5]
    \mpsWire{0}{-1.75}{0}{0.75}
    \gridLine{-0.75}{0}{0.75}{0}
    \gridLine{-0.75}{-1}{0.75}{-1}
    \mpsC{0}{-1}{0}{colMPSBetaB}
    \wedgeL{0}{0}
    \wedgeL{0}{-1}
    \node at (-1.25,0) {$1$};
    \node at (1.25,0) {$1$};
    \node at (-1.25,-1) {$1$};
    \node at (1.25,-1) {$1$};
  \end{tikzpicture}=
  \begin{bmatrix}
    \Pi_{j-1} \vartheta & 0 & 0 \\
    \Pi_{j-1} (1-\vartheta) & 0 & 0 \\
    0 & 0 & 0
  \end{bmatrix}.
\end{aligned}
\end{equation}

\section{Details about the numerical evaluations}\label{appC}
\subsection{Free fermions}
\begin{figure}
  \centering
  \begin{subfigure}[c]{0.48\textwidth}
    \centering
    \includegraphics[width=\linewidth]{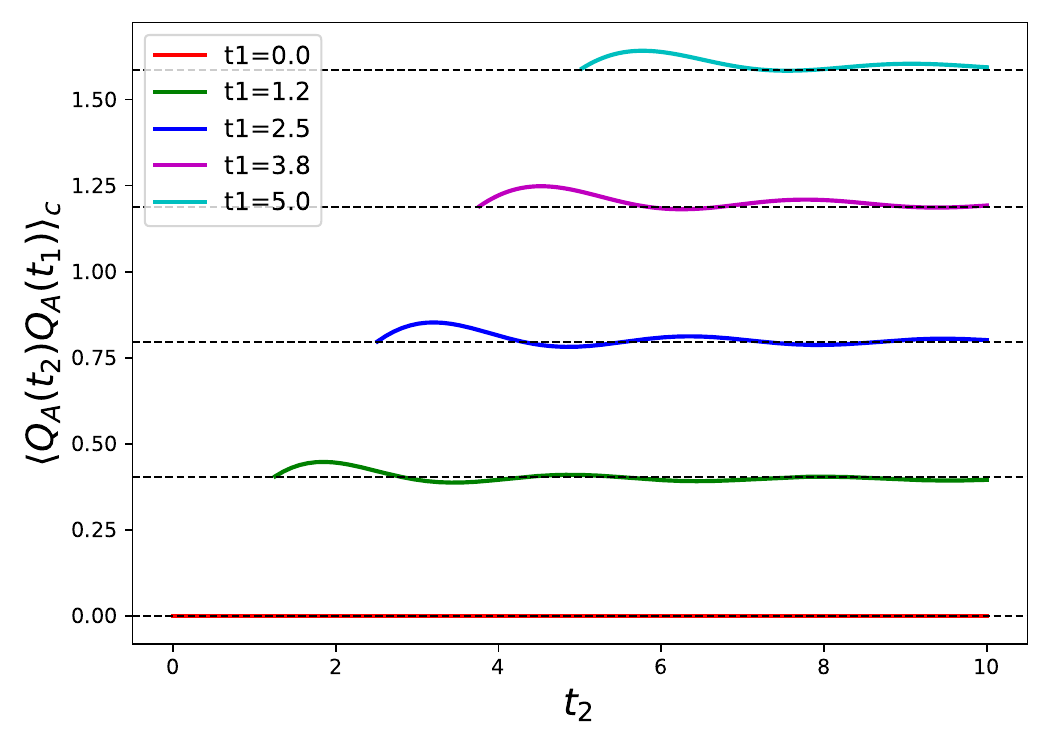}
  \end{subfigure}
  \hfill
  \begin{subfigure}[c]{0.48\textwidth}
    \centering
    \includegraphics[width=\linewidth]{figures/free_correlator_L_A=40.pdf}
  \end{subfigure}
  \caption{\label{fig:freeplots1}
  Connected two-point function, for various values of $t_1$, obtained by varying $t_2$. The two plots are performed at different values of $\ell_A$, which is set to $20$ in the left and to $\ell_A = 40$ in the right. Although the system sizes are relatively small, the oscillations about the asymptotic prediction (dashed lines) are already small, and there appears to be no deviation from it.}
\end{figure}

\begin{figure}
    \begin{subfigure}[c]{0.48\textwidth}
        \centering
        \includegraphics[width=\linewidth]{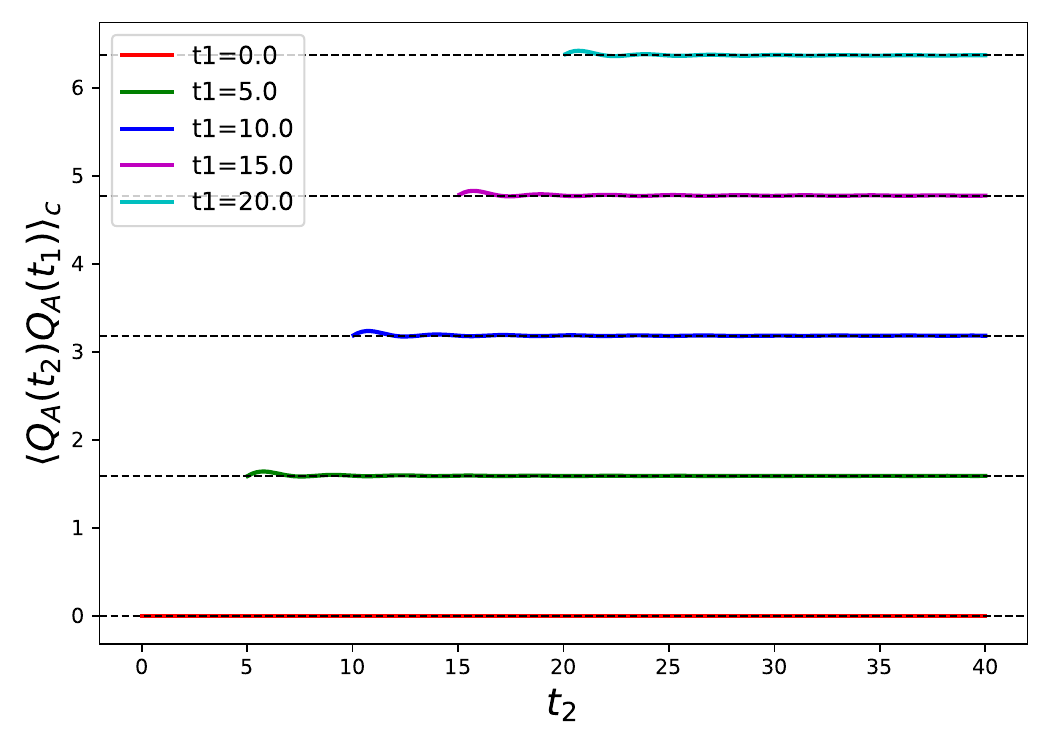}
    \end{subfigure}
    \hfill
    \begin{subfigure}[c]{0.48\textwidth}
        \centering
        \includegraphics[width=\linewidth]{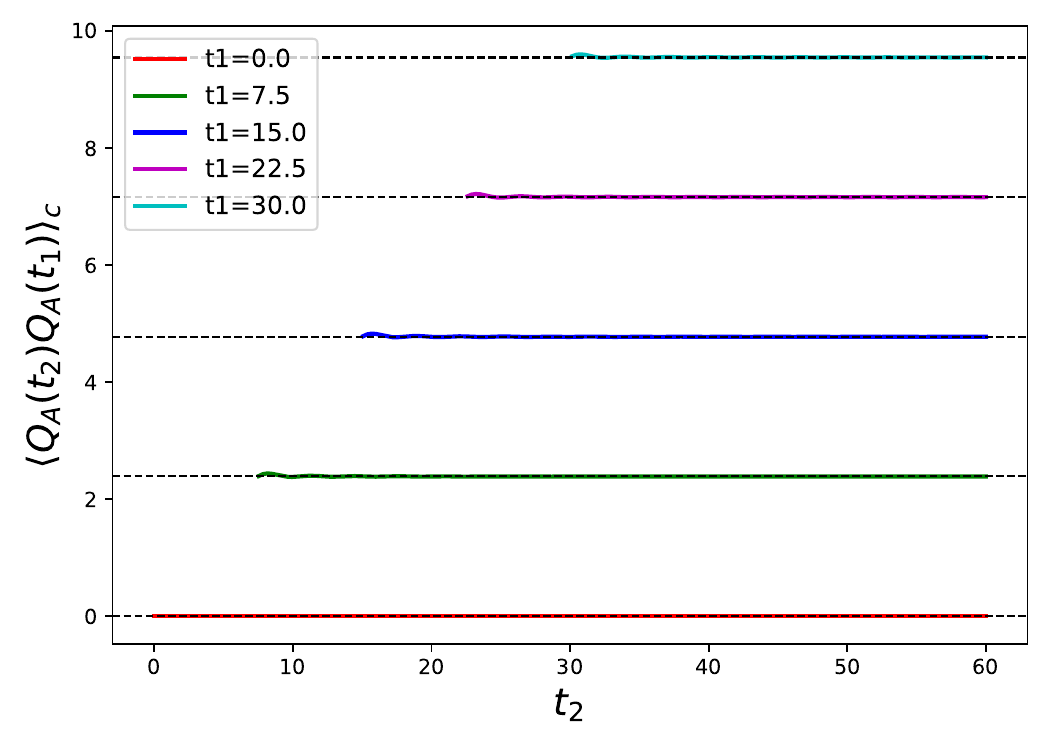}
    \end{subfigure}
    \caption{\label{fig:freeplots2} Same as the previous figure, with larger values of $\ell_A$, which is set to 80 in the left plot and 120 in the right. For these values, the deviations of the correlator from the asymptotic prediction are negligible, showing that the analytical prediction is asymptotically exact.}
\end{figure}

In free fermionic systems evolving from Gaussian states, any time-ordered correlation function of the charge can be evaluated from the correlation matrices $C_{xy}(t_2,t_1) = \expval*{c_x^\dag(t_2) c_y(t_1)}$ and $\tilde C_{xy}(t_2,t_1) = \expval*{c_x(t_2) c^\dag_y(t_1)}$ and the  by direct application of Wick theorem. The analytical expression for the correlation matrix can be obtained easily in momentum space,
\begin{equation} \label{eq:corrmatdef}
  \expval{c_x^\dag(t_2) c_y(t_1)}=\frac{1}{L}\sum_{k,q} 
  \mathrm{e}^{ikx - i qy} \mathrm{e}^{i\varepsilon_k t_1 - i \varepsilon_q t_2} \expval{c_k^\dag c_q}.
\end{equation}
For simple initial states, the correlation function $\expval*{c_k^\dag c_q}$ has simple expressions: focussing on the Néel state as a prototypical example, this is $\expval*{c_k^\dag c_q} = \frac{1}{2} (\delta_{k,q} + \delta_{k,q-\pi})$. Taking the thermodynamic limit of \eqref{eq:corrmatdef}, and observing that the energies in the hopping model are simply $\varepsilon_k = - \cos k$, the sum can be written simply in terms of Bessel functions owing to the identity 
\begin{equation}
  \int_0^{2\pi}\frac{{\rm d}k}{2\pi} \mathrm{e}^{i k z} \mathrm{e}^{it\cos k} = i^z J_z(t),
\end{equation}
leading to the final result 
\begin{equation}
  \begin{aligned}
    C_{xy}(t_2,t_1) &= \frac{1}{2}i^{y-x}
    \Big[J_{x-y}(t_2-t_1) + (-1)^y J_{x-y}(t_2+t_1)\Big], \\
    \tilde C_{xy}(t_2,t_1) &= \frac{1}{2}i^{x-y}
    \Big[J_{x-y}(t_2-t_1) - (-1)^y J_{x-y}(t_2+t_1)\Big].
  \end{aligned}
\end{equation}
We can then proceed to the evaluation of the correlation functions of the charge: considering for example the two point function, this can be expanded as
\begin{equation}
  \begin{aligned}
    \expval{Q_A(t_2)Q_A(t_1)} &= \sum_{x,y \in A} \expval{n_x(t_2) n_y(t_1)} 
    =  \sum_{x,y \in A} \Big[\expval*{n_x(t_2)}\!\expval*{n_y(t_1)} 
    + \expval*{c_x^\dag (t_2) c_y(t_1)}\!\expval*{c_x(t_2) c^\dag_y(t_1)}\Big] \\
    &= \expval{Q_A(t_2)}\expval{Q_A(t_1)} 
    +\sum_{x,y \in A} \expval*{c_x^\dag (t_2) c_y(t_1)}\! \expval*{c_x(t_2) c^\dag_y(t_1)},
  \end{aligned}
\end{equation}
where the first term simplifies when considering the connected correlator, giving (for $t_2>t_1$)
\begin{equation}
  \expval{Q_A(t_2)Q_A(t_1)}_c = \sum_{x,y \in A} 
  \expval*{c_x^\dag (t_2) c_y(t_1)}\!\expval*{c_x(t_2) c^\dag_y(t_1)} 
  = \operatorname{Tr}[C(t_2,t_1) \tilde C^T(t_2,t_1)].
\end{equation}
This can be computed efficiently, allowing us to consider a wide range of values of $\ell_A$, $t_1$ and $t_2$. In particular, from the plots in Figs.~\ref{fig:freeplots1} and \ref{fig:freeplots2} it is evident that the correlator only depends on $t_1$ up to small fluctuations even for small system sizes. This is not a completely unexpected feature in free fermionic theories, where in general results obtained through the quasiparticle picture emerge at system size which are much smaller than those needed in the interacting case. In this context, this can be related to the fact that correlations in time in the free case are much smaller than in the integrable interacting case.

\begin{figure}[]
    \begin{subfigure}[c]{0.48\textwidth}
        \centering
        \includegraphics[width=\linewidth]{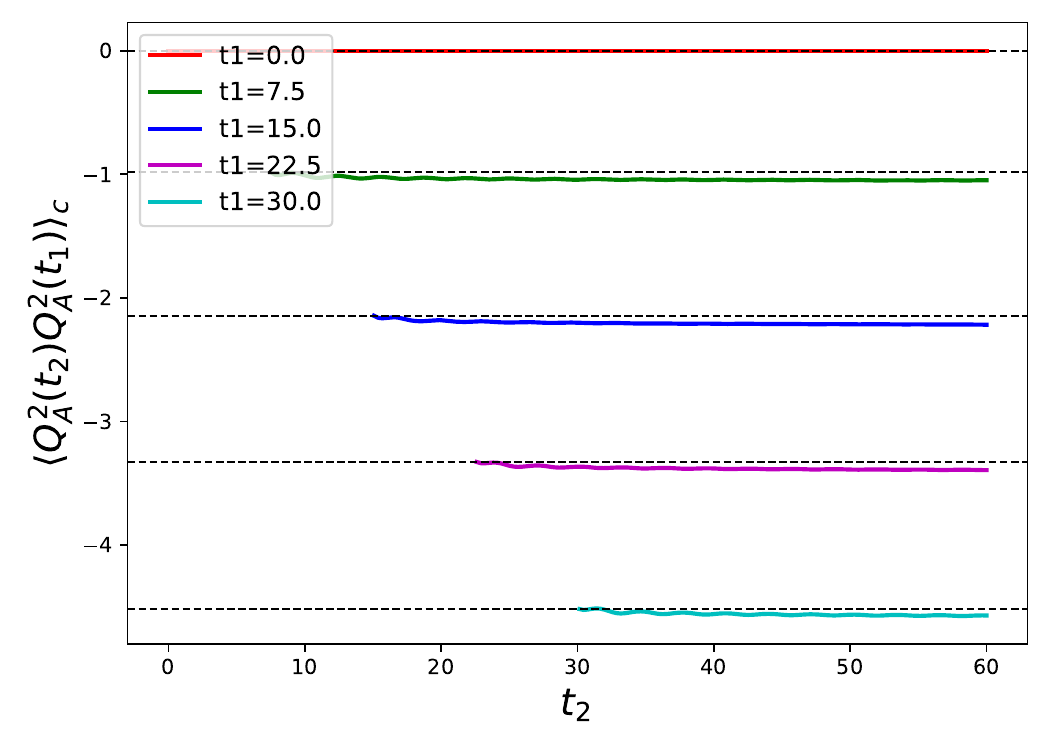}
    \end{subfigure}
    \hfill
    \begin{subfigure}[c]{0.48\textwidth}
        \centering
        \includegraphics[width=\linewidth]{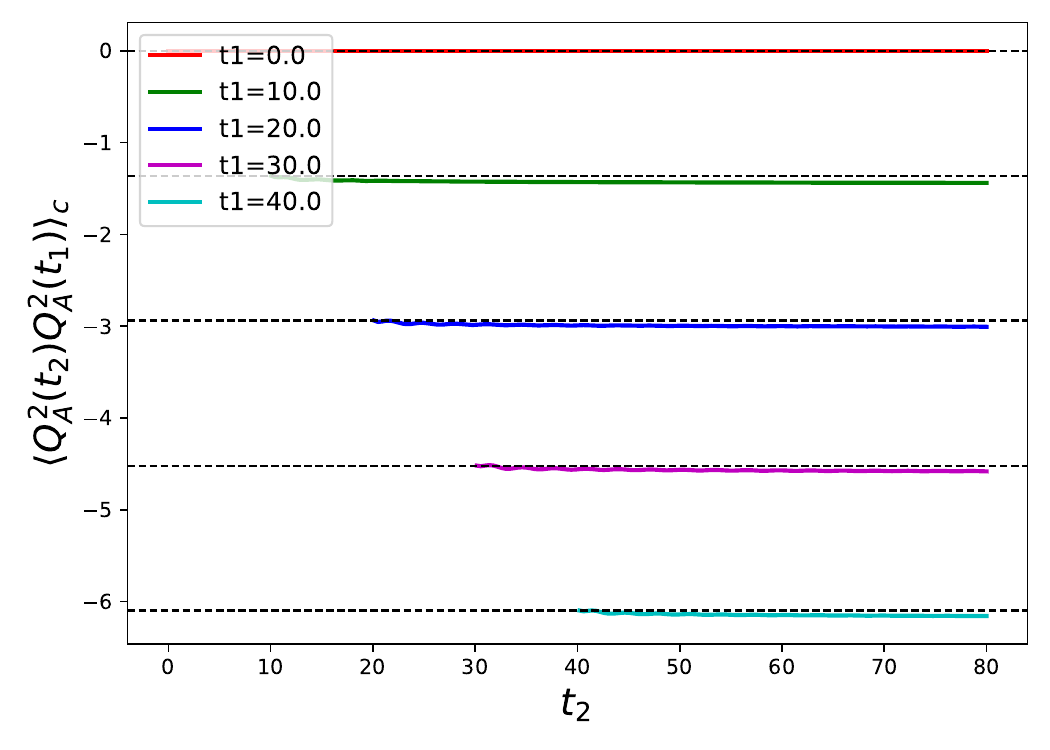}
    \end{subfigure}
    \caption{\label{fig:freeplots4point} Connected four point function $ \expval*{Q^2_A(t_2) Q^2_A(t_1)}_c $, in which the times are pairwise taken the same. System size is taken to be $\ell_A=120$ on the left and $\ell_A=160$ on the right: the small subleading correction to the asymptotic result is indeed washed away increasing system size and value of $t_1$, as claimed in the main text.}
\end{figure}

For three and higher point functions, the discussion is completely analogous, as everything can be similarly reduced to the two point correlation matrices. In general the connected correlator is obtained by summing over permutations of elements which are made of full cycles, as all other terms are cancelled when removing the disconnected contributions. For the three point function there are only two such cycles, giving
\begin{equation}
    \expval{Q_A(t_3) Q_A(t_2) Q_A(t_1)}_c = \operatorname{Tr}[C(t_3,t_1) \tilde C^T(t_2,t_1) \tilde C^T(t_3,t_2)] - \operatorname{Tr}[C(t_3,t_2)  C(t_2,t_1) \tilde C^T(t_3,t_1)] .
\end{equation}
For the four point function $\expval*{Q_A(t'_2) Q_A(t_2) Q_A(t'_1) Q_A(t_1)}_c$, 6 terms would be present in general. The situation is simplified by  restricting to two times, $t'_1=t_1$ and $t'_2=t_2$; in this case the terms combine to give
\begin{equation}
    \expval*{Q^2_A(t_2) Q^2_A(t_1)}_c = \operatorname{Tr}\left[ \Gamma(t_2)C(t_2,t_1)\Gamma(t_1) \tilde C^T(t_2,t_1)\right] - 2 \operatorname{Tr}\left[(C(t_2, t_1) \tilde C^T(t_2, t_1))^2\right]
\end{equation}
where $\Gamma(t_i) =\left(2 C(t_i,t_i)- 1\right) $. The quench behaviour of the four point function is shown in Fig.~\ref{fig:freeplots4point} for larger system sizes compared to the main text. Clearly the curves exhibit the main feature expected by our prediction, namely the independence of the result on $t_2$. Compared to the two point function, the result exhibits a slight offset with respect to the value at $t_1$. This is not too surprising, as our results refer to the leading order, and subleading corrections are expected to arise; it is in fact more surprising that no such subleading contribution is present in the 2-point function. The plots, compared to the one shown in the main text, show clearly that the subleading contribution is increasingly less relevant as the system size is increased.

\subsection{Interacting theories}
In the interacting case, we perform a tensor network evaluation using the iTensor library \cite{iTensor}.
By representing the initial state as a matrix product state the time evolution can be implemented approximately by performing a Trotter-Suzuki decomposition of the time evolution operator, 
\begin{equation}
  {{U}}(\delta t) = \prod_{i=1}^{L-1} \mathrm{e}^{-ih_{j,j+1} \delta t}  
  \prod_{j=1}^{L-1} \mathrm{e}^{-ih_{L-j,L-j+1}\delta t}
\end{equation}
where each term $\mathrm{e}^{-ih_{i,i+1} \delta t} $ can be viewed as a quantum gate acting on two spins. This produces errors of order $\delta t^3$ per time step, and allows to obtain good numerical results for $\delta t = 0.01-0.1$. In order to avoid excessive growth of the bond dimension, we perform a mixed Schrodinger-Heisenberg evolution, by computing
\begin{equation}
  \mel{\psi(t_1)}{\mathrm{e}^{i\beta_2 Q_A(t_2-t_1)} \mathrm{e}^{i\beta_1 Q_A}}{\psi(t_1)}.
\end{equation}
This is convenient since $Q_A$ is an operator which commutes with the Hamiltonian in the bulk, and therefore its Heisenberg evolution is concentrated at the boundaries of $A$ and the growth of bond dimension is expected to remain reasonably small.
This allows to keep the maximum bond dimension to $\chi_s^{\rm max} = 2048$ for the state evolution and $\chi_o^{\rm max} = 512$ for the operator evolution. The main numerical limitation comes from the evaluation of the correlator itself, since the cost evaluation of the two point function scales as $O(L\chi_s^3\chi_o^2)$. With our choice of evolution, we are able to reach maximum final times $t_2 =18$ keeping good precision.

\begin{figure}[ht!]
    \begin{subfigure}[c]{0.48\textwidth}
        \centering
        \includegraphics[width=\linewidth]{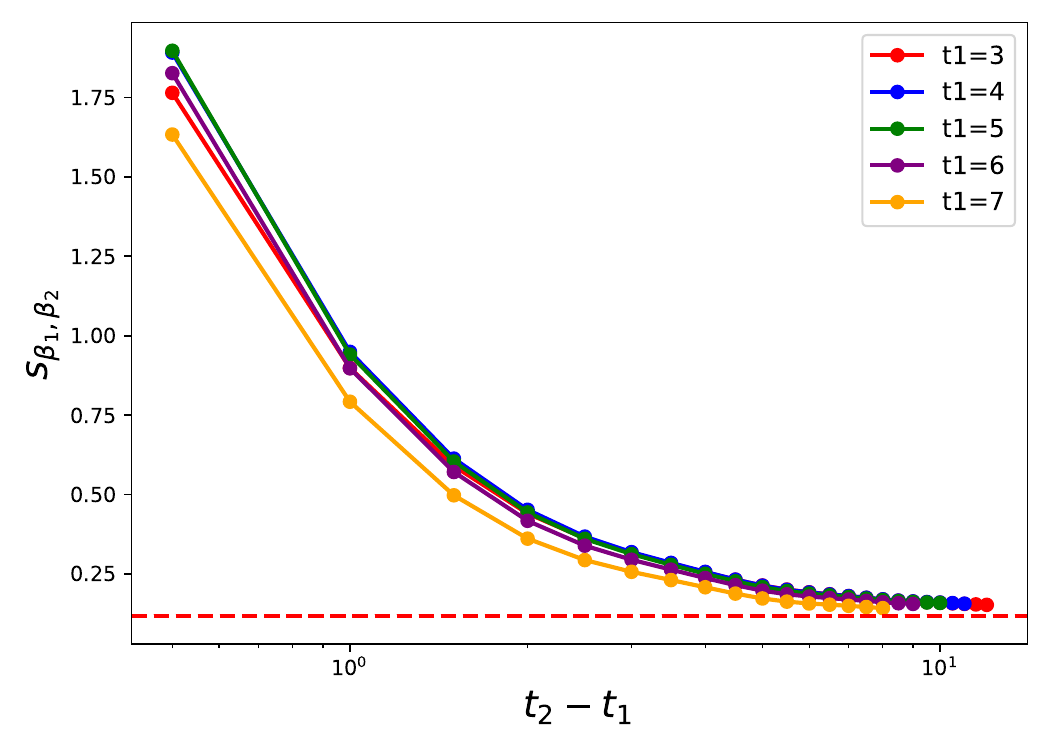}
    \end{subfigure}
    \hfill
    \begin{subfigure}[c]{0.47\textwidth}
        \centering
        \includegraphics[width=\linewidth]{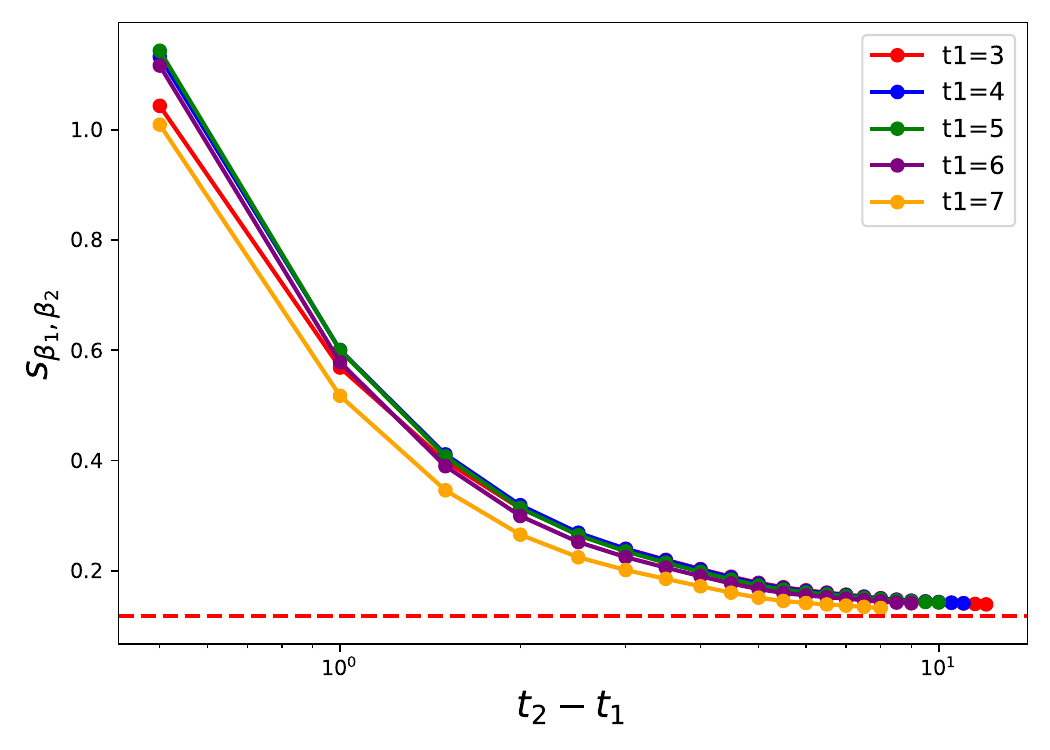}
    \end{subfigure}
    \caption{\label{fig:FCSpi8pi4}
      Time dependence of $s_{\beta_1,\beta_2}$ for several values of $t_1$ for a subsystem  of size $\ell_A=40$. The plot on the left represents the choice of parameters $\beta_1 = \pi/4$, $\beta_2 = \pi/4$, while the second one $\beta_1 = \pi/8$, $\beta_2 = \pi/4$. Although the values at finite times differ, the two plots show that the asymptotic result obtained is the same, as would be expected since the two plots share the same value of $\beta_2=\pi/4$.  }
\end{figure}

\begin{figure}[ht!]
    \begin{subfigure}[c]{0.48\textwidth}
        \centering
        \includegraphics[width=\linewidth]{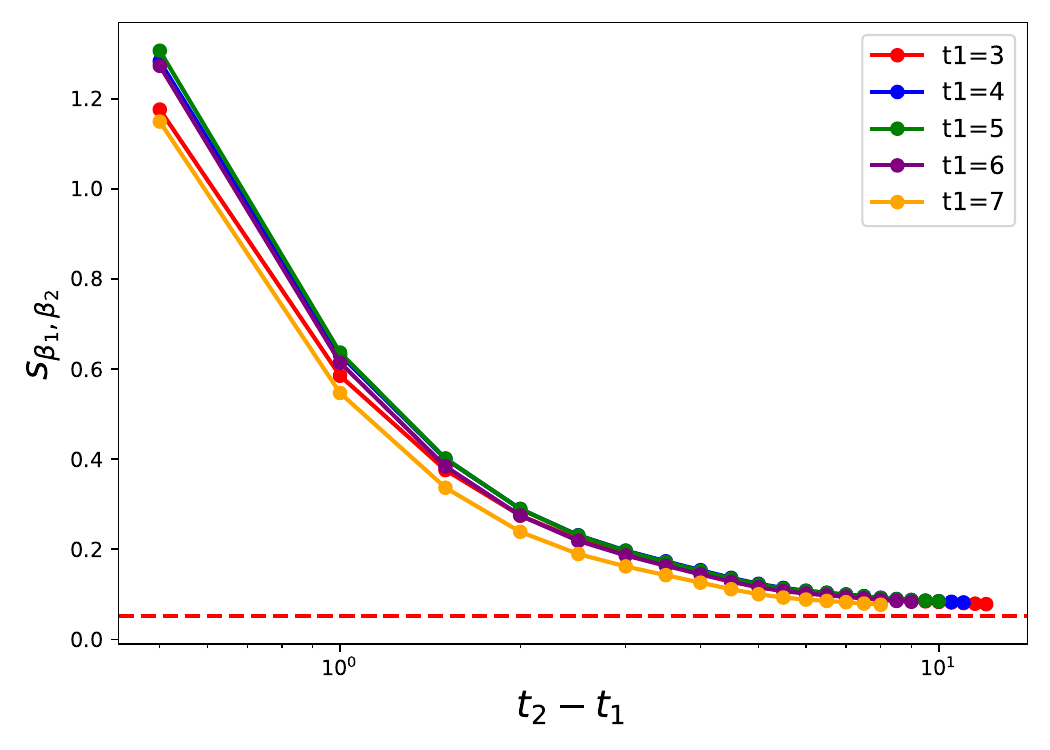}
    \end{subfigure}
    \hfill
    \begin{subfigure}[c]{0.47\textwidth}
        \centering
        \includegraphics[width=\linewidth]{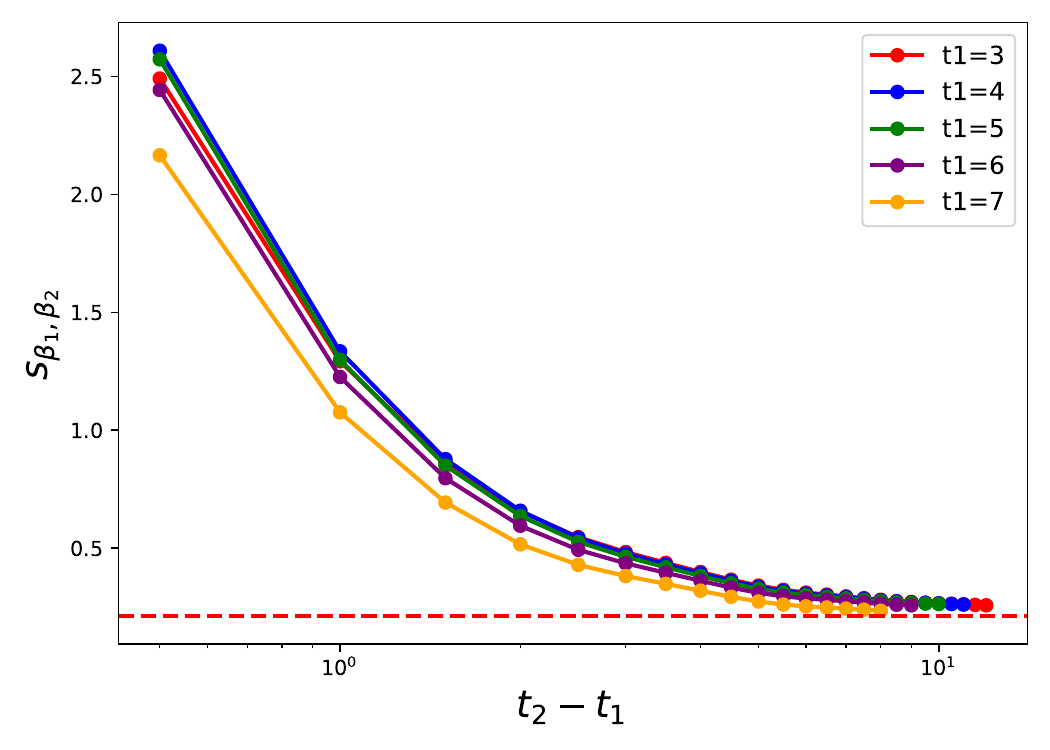}
    \end{subfigure}
    \caption{\label{fig:FCSpi4pi3} Time dependence of $s_{\beta_1,\beta_2}$ for several values of $t_1$ for a subsystem  of size $\ell_A=40$; the parameters are set to $\beta_1 = \pi/4$, $\beta_2 = \pi/6$ on the left, and to $\beta_1 = \pi/4$, $\beta_2 = \pi/3$ on the right. In this case, the two plots share the same value of $\beta_1$, but the asymptotic value is different because of the difference in the values of $\beta_2$.}
\end{figure}

As discussed in the main text, the theoretical expression of interest is 
\begin{equation}
  \mathcal{F}^{(2)}_A(\{t_i,\beta_i\})= 
  t_1\Gamma\left(\beta_1 + \beta_2\right) + (t_2-t_1)\Gamma\left( \beta_{2}\right).
\end{equation}
The main feature of this expression is that, at fixed $t_1$, the evolution is predicted to be linear in $t_2-t_1$ with slope $\Gamma\left( \beta_{2}\right)$; or, equivalently, that there is a time-shell structure. Therefore it is natural to consider the slope 
\begin{equation}
    s_{\beta_1,\beta_2} = \frac{1}{t_2-t_1} \left(\mathcal{F}^{(2)}_A(\{t_i,\beta_i\})- t_1\Gamma\left(\beta_1 + \beta_2\right)\right),
\end{equation}
such that $s_{\beta_1,\beta_2} \xrightarrow{\Delta t\to \infty}\Gamma\left( \beta_{2}\right) $ where our theory predicts this result to hold asymptotically in the time separation.
By numerically evaluating $s_{\beta_1,\beta_2}$, we are simultaneously confirming two things: first of all, that the dependence on $t_1$ alone is fully removed by subtracting the second term. Secondly, that the residual term is precisely linear in $\Delta t$, with a slope which is fixed by the result for the $1-$FCS. These features are clearly confirmed by Fig. \ref{fig:FCSpi8pi4} and Fig. \ref{fig:FCSpi4pi3}, which show $s_{\beta_1,\beta_2}$ for several values of $t_1$, and several values of the parameters $\beta_1$ and $\beta_2$.

\begin{figure}
    \begin{subfigure}[c]{0.48\textwidth}
        \centering
        \includegraphics[width=\linewidth]{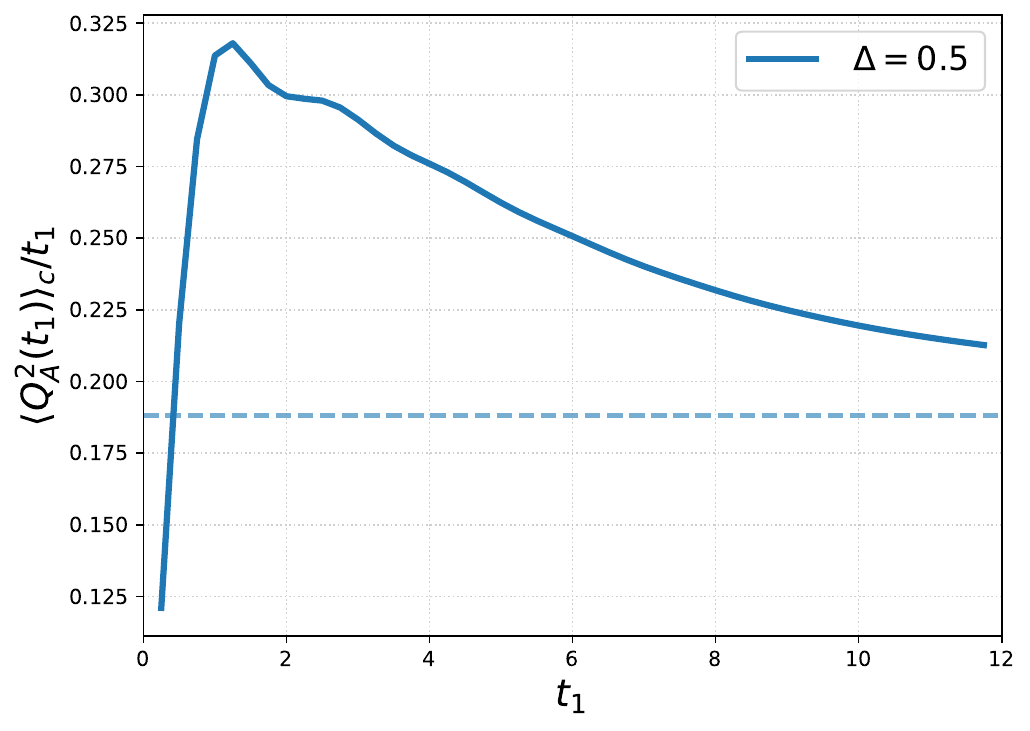}
    \end{subfigure}
    \hfill
    \begin{subfigure}[c]{0.47\textwidth}
        \centering
        \includegraphics[width=\linewidth]{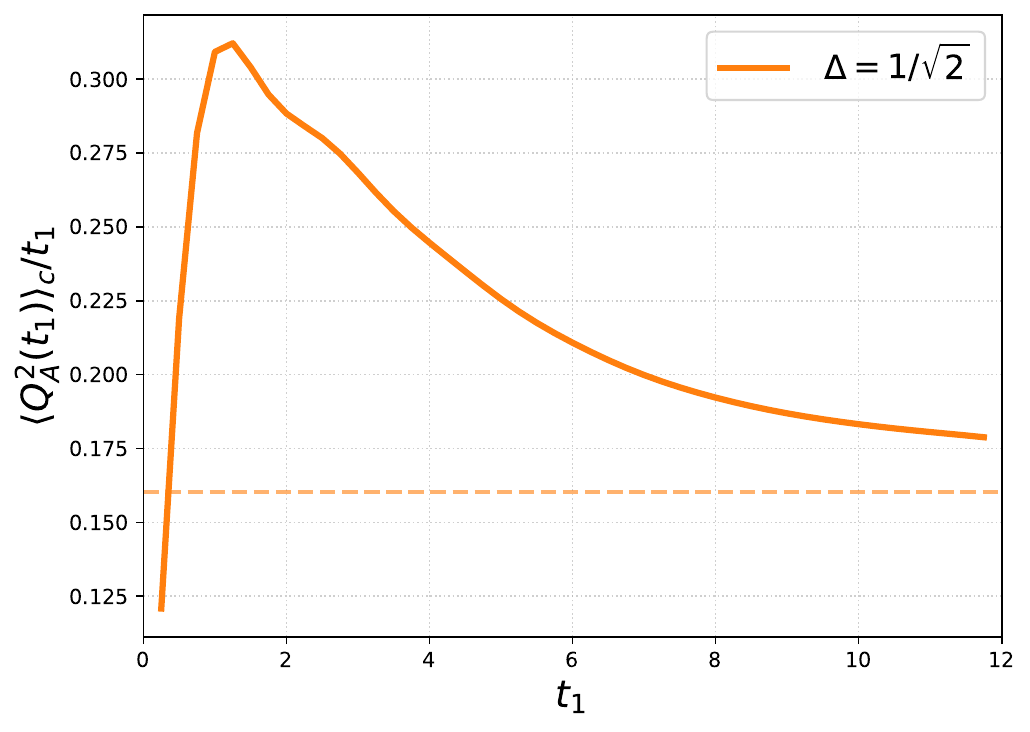}
    \end{subfigure}
    \caption{Time dependence of the connected two point function at a single time for two values of $\Delta$, in a quench from the Néel state. In both cases, the approach of the one point function to the asymptotic value is slow, and shows significant divergence at the time scale which can be reached through TEBD. }
    \label{fig:2cumulant}
\end{figure}

Regarding the behaviour of the correlation functions, the first observation is that the approach of $\expval{Q_A^2(t_1)}_c$ to its asymptotic prediction \eqref{eq:asymptoticcumulant} is rather slow, as shown in Fig.~\ref{fig:2cumulant}. While in the free fermionic case the prediction for $\expval{Q_A(t_2)Q_A(t_1)}_c$ was compared to the instantaneous value of $\expval{Q_A^2(t_1)}_c$, this could be done only because in the free fermionic case the approach to the asymptotic value takes place on shorter time scales. If this does not hold, the two point function has to be compared with the asymptotic prediction, which is expected to reproduce good results if both $t_1$ and $t_2-t_1$ are sufficiently large.
\end{document}